# Frictional adhesive contact of multiferroic coatings based on the hybrid element method

Yanxin Li [1], Bo Pan [3], Yun Tian [3], Lili Ma [4], Nicola Menga [2], Xin Zhang [1, *]

1. School of Mechanical and Electrical Engineering, University of Electronic Science and Technology of China, Chengdu 611731, China
2. Department of Mechanics, Mathematics and Management, Polytechnic University of Bari, Via Orabona 4, 70125 Bari, Italy
3. Jiaxing Guodiantong New Energy Technology Co., Ltd, Jiaxing 314001, China
4. School of Mathematics and Statistics, Ningxia University, Yinchuan 750021, China

* Corresponding authors: Xin Zhang, zhangxin@uestc.edu.cn

**Abstract**

We study the frictional adhesive contact of a rigid insulating sphere sliding past a multiferroic coating deposed onto a rigid substrate, based on the hybrid element method (HEM). The adhesion behavior is described based on the Maugis-Dugdale (MD) model. The adhesion-driven conjugate gradient method (AD-CGM) is employed to calculate the distribution of unknown pressures, while the discrete convolution-fast Fourier transform (DC-FFT) is utilized to compute the deformations, surface electric and magnetic potentials as well as the subsurface stresses, electric displacements, and magnetic inductions. We found that the coating thickness affect the contact stiffness and the interplay between friction and adhesion. More importantly, friction and gap-dependent MD adhesion affects elastic, electric, and magnetic behavior of the interface, breaking the symmetry between leading and trailing edges behaviors in all the investigated fields. Indeed, increasing the friction coefficient, the contact shape is no longer circular, the pressure distribution shifts towards the leading edge, the electric/magnetic surface potentials distributions sharpen at the leading edge, and the subsurface stress fields concentrates at the trailing edges.

# 1. Introduction

Multiferroic composites consist of two or more ferroic orders, derived from components exhibiting properties like ferromagnetism, antiferromagnetism, ferroelectricity, or ferroelasticity [1]. These composites, commonly utilized as surface coatings, hold significant potentials as intermediate layers in the highly integrated electronic components, power batteries and energy transducers [1–3]. In all these applications, multiferroic interfaces usually experience adhesion and friction, which eventually exert a substantial influence on critical factors such as the magnetoelectric coupling effect, surface physics, as well as contact and fracture mechanics [4,5]. Therefore, understanding the frictional adhesive contact behavior of these interfaces depending on the coating geometry (thickness) and electric/magnetic surface charge densities is pivotal to improve the design and facilitate the production of multiferroic coatings.

Early investigations into adhesion in contact mechanics were conducted by Bradley, who examined the adhesion contact of rigid spheres [6]. Subsequently, two widely recognized contact theories emerged able to take into account for the spheres elasticity: the Johnson-Kendall-Roberts (JKR) [7] theory, and the Derjaguin-Muller-Toporov (DMT) [8] one. The former is applicable relatively soft contacts with a larger area and postulate infinitely short-range adhesive interactions, while the latter is better suited for hard contacts and considers long-range attractive forces. Later on, based on the superposition of Hertzian solution for spherical contacts and Dugdale solution for cracks under internal loading, Maugis [9] derived a closed form solution for adhesive contacts, namely the Maugis-Dugdale (MD) model, able to smoothly represent the JKR-DMT transition as a function of a single parameter $\lambda$ (Maugis parameter). By exploiting a similar superposition procedure, analytical solutions for the magneto-electro-elastic contact problem of a rigid spherical [10] and conical [11] punches against multiferroic composite half-space have been derived in the JKR, MD, and DMT adhesive regimes. Similarly, for rigid indenters in contact with piezoelectric materials, Chen et al. [12] obtained the general solutions in the JKR and MD adhesive regimes using the Hankel transformation and the superposition principle. Utilizing the superposition principle and the equivalent indentation method, Jin and Guo [13,14] generalized the study to the case of adhesive contacts involving an axisymmetric rigid punch of arbitrary shapes and an elastic half-space. In addition to the aforementioned studies, linear superposition is also used by Li and Liu [15,16] to tackle the case of axisymmetric contacts with multi-layer elastic solids in MD adhesive regime, ultimately deriving the closing condition as a singular integral equation by using the zero-order Hankel transformation. Stan et al. [17] derived analytical solutions for the stress field and electric displacement field in the adhesive contact using the transfer matrix method, achieved through the resolution of coupled singular integral equations. Rey et al. [18] investigated the adhesive contact between half-spaces with rough surfaces using the FFT-based boundary element method (BEM). Sergici et al. [19] investigated the frictionless adhesive contact behavior between a spherical indenter and an elastic-layered medium based on the MD model. Bazrafshan et al. [20] utilized the adhesion stress framework outlined in the MD model. They introduced an extended conjugate gradient method (CGM) to explore the adhesive contact behavior of a rigid sphere against a wavy elastic half-space with different aspect ratios. Their findings indicated that CGM is well-suited for analyzing frictionless adhesive contact interactions between two elastic bodies characterized by intricate surface geometries.

Usually, when two contacting bodies slide past each other, friction arises, opposing the relative motion, which can have different chemo-physical origins. For instance, in the presence of viscoelastic materials, rough rigid surfaces in sliding motion induce cyclic deformations in the bulk of the viscoelastic solid, thus entailing energy dissipation and friction. This mechanism has been widely investigated by means of mean-field theories [21,22] and deterministic calculations [23–25]. In the presence of adhesion, the relative sliding motion between the deformable solid and the rigid indenter leads to the propagation of cracks at the edges of each contact spot: closing one at the leading edges, and an opening one at the trailing edges. In this context, even in purely elastic materials, adhesion may induce an additional mechanism for localized energy dissipation near the contact edges, which is usually associated to small-scale viscoelasticity and local nonequilibrium interfacial processes. Consequently, leading and trailing edges may present different adhesive behaviors (*i.e.*, adhesion hysteresis), eventually leading to friction opposing the sliding motion. Barquins [26] firstly quantified the frictional response of rigid cylinders rolling over flat smooth surfaces of natural rubber, while She et al. [27] later showed that adhesion hysteresis and friction can be related to the process of adsorption and desorption of polymer chains on a PDMS flat surface, so that the energy required to open a crack at the contact trailing edge is markedly greater than that needed to close it at the leading one. Later on, Ghatak *et al*. [28] investigated the sliding contact behavior of elastomeric polymers on low energy surfaces and established a connection between molecular rate processes and the observed values of interfacial adhesion and friction. Carbone *et al*. [29] focused on a purely elastic wavy contact, postulating small-scale viscoelasticity localized close to the contact edges. Consequently, the cracks experience different energy release rates at the advancing and receding edges (*i.e.*, adhesion hysteresis), leading to the emergence of friction and asymmetric contact shape. Hao *et al*. [30] explored the influence of surface adhesion and friction in steady-state rolling contacts by employing a hybrid superposition of indentation and rolling contacts solutions, also quantifying the resistance to rolling motion (*i.e.*, adhesion friction coefficient). A contact model based on Lennard-Jones interfacial potentials for rigid spheres and micro-structured surfaces was developed by Zhang *et al*. [31]; while the same authors later focused on the rolling torque in a ball on flat contact [32] arising from an (empirically) given adhesion hysteresis, which is modeled by heterogeneous surface potentials. Recently, Carbone *et al*. [33,34] developed a theory for adhesive sliding contacts in the framework of viscoelastic materials. The found that the interaction between adhesion and viscoelasticity leads distinct energy dissipation rates at the contact trailing and leading edges thus leading to adhesion hysteresis. The overall friction opposing the sliding motion and the interface behavior depend on both bulk viscoelastic dissipation and adhesion hysteresis.

Friction may also exist independently of viscoelasticity and adhesion, as phenomenologically modeled firstly by Amontons/Coulomb, and then generalized by Derjaguin [34] and Bowden and Tabor [35]. Even in this case, friction significantly affects the contact response of a solid interface, especially when thin coating exists, as in the case of multiferroic composites. In a series of recent papers on line [36,37] and areal [38] contacts, Menga and co-authors have shown that, regardless of their origin, in-plane shear stresses impinge on interfacial key quantities, such as contact area size and shape, surface stress concentration, gap distribution, and leakage, thus entailing non-negligible impacts on component functionality such as electrical conductivity in solid state batteries [39] and leak flow rate in static and rotary seals. This depends on the elastic coupling between in-plane and out-of-plane displacements and stresses prescribed in linear elasticity for compressible and/or thin solids, such as multiferroic coatings and functionally graded materials [40].

The above studies highlighted the key role of the adhesion and friction on the contact responses within the elastic or viscoelastic framework, also with reference to wear of sliding materials [63]. For the multiferroic coatings, the coupling effects of adhesion and friction on the contact behaviors are expected to be more complex and critical, due to the additional magneto-electro-elastic coupling.

The hybrid element method (HEM) is different from the full numerical method of the finite element method (FEM), as the former combines the theoretical basic solutions with the fast numerical algorithm [41]. This paper investigates of the magneto-electro-elastic contact problem of a rigid sphere sliding past a multiferroic coating, in the presence of interfacial adhesion and friction. The problem is modeled with HEM, and calculations rely on the discrete convolution fast Fourier transform (DC-FFT) and the extended conjugate gradient method (CGM). The coupled effects of friction, adhesion, and electromagnetic field on the pressure distribution, electric potential distribution, magnetic potential distribution, and subsurface stresses are analyzed in detail.

## 2. Frictional adhesive contact of multiferroic coatings

### 2.1 Problem description of the frictional adhesive contact

**Figure 1(a)** depicts a sliding contact model between a sphere and a coating-substrate system under a constant normal load $P$ and a tangential load $Q$. The sphere, with a radius of $R$, is rigid, electric/magnetic insulating, while the coating with thickness $h_t$ is made of transversely isotropic multiferroic materials, perfectly bonded to a rigid substrate. A Cartesian coordinate system $(x,y,z)$ is introduced with the origin at the contact center, where the $x-y$ plane is parallel to the transversely isotropic plane of the coating. Both contact bodies are free of body forces, as well as volumetric electric and magnetic charges.

The contact surface is illustrated in **Figure 1**(b), considering the magneto-electro-mechanical coupling, surface friction, and adhesion effects. We assume a Maugis-Dugdale (MD) adhesion [9] at the interface; therefore, the inner contact region of radius $a$ (where the contact gap $g=0$) is surrounded by an annular region (with an inner radius $a$ and an outer radius $c$) where uniform cohesive tractions $\sigma_0$ occurs for $g \leq h_0$, where $h_0$ is the maximum adhesion distance. Following the Dugdale approximation of interatomic potentials, the resulting adhesion energy is $\Delta\gamma_0 = \sigma_0 h_0$ (see **Figure 1**(c)).

Multiferroic contacts involve stiff materials which may experience relatively high loads with a real surface morphology far from being atomically smooth; therefore, following Homola et al. [42], we neglect adhesive contribution to friction and assume Coulomb friction t the interface, so that the shear stresses are given by $p_x = \mu p_z$ where repulsive interactions occurs, i.e., where $p_z \geq 0$, with $p_z$ being the normal pressure and $\mu$ the friction coefficient. In the contact region, surface electric and magnetic charge densities $q_b$ and $g_b$, respectively, are prescribed.

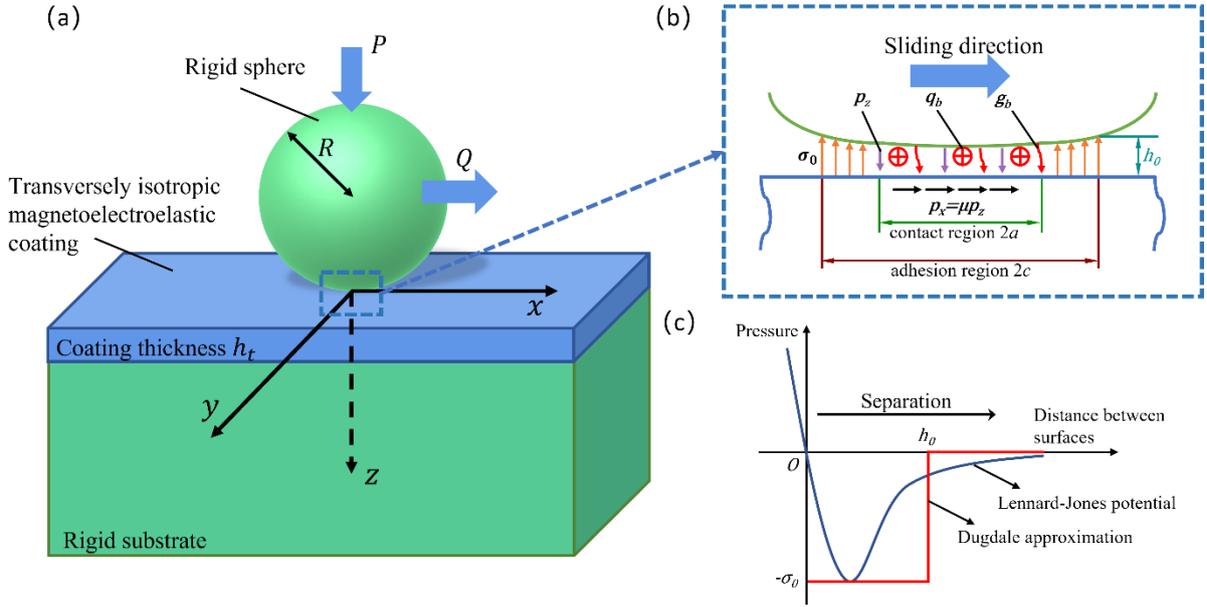

**Figure 1** (a) Illustration of the three-dimensional frictional adhesive contact of a rigid and insulating sphere sliding past a transversely isotropic multiferroic coating deposed onto a-rigid substrate. The spere of radius of $R$ is subjected to a normal load $P$ and a tangential load $Q$. (b) The schematics of electric and magnetic surface charges, as well as mechanical tractions acting on the multiferroic coating surface. (c) The Dugdale cohesive approximation of the Lennard-Jones potential.

## 2.2 Basic equations and boundary conditions

Following previous studies on multiferroics [43–45], the constitutive equations for a transversely isotropic multiferroic material can be expressed as follows:

$$\sigma_{xx} = c_{11}u_{x,x} + c_{12}u_{y,y} + c_{13}u_{z,z} + e_{31}\phi_{,z} + d_{31}\varphi_{,z} \tag{1}$$

$$\sigma_{yy} = c_{12}u_{x,x} + c_{11}u_{y,y} + c_{13}u_{z,z} + e_{31}\phi_{,z} + d_{31}\varphi_{,z} \tag{2}$$

$$\sigma_{zz} = c_{13}u_{x,x} + c_{13}u_{y,y} + c_{33}u_{z,z} + e_{33}\phi_{,z} + d_{33}\varphi_{,z} \tag{3}$$

$$\sigma_{zy} = c_{44}u_{y,z} + c_{44}u_{z,y} + e_{15}\phi_{,y} + d_{15}\varphi_{,y} \tag{4}$$

$$\sigma_{zx} = c_{44}u_{x,z} + c_{44}u_{z,x} + e_{15}\phi_{,x} + d_{15}\varphi_{,x} \tag{5}$$

$$\sigma_{xy} = c_{66}u_{x,y} + c_{66}u_{y,x} \tag{6}$$

$$D_x = e_{15}u_{x,z} + e_{15}u_{z,x} - \varepsilon_{11}\phi_{,x} - g_{11}\varphi_{,x} \tag{7}$$

$$D_y = e_{15}u_{y,z} + e_{15}u_{z,y} - \varepsilon_{11}\phi_{,y} - g_{11}\varphi_{,y} \tag{8}$$

$$D_z = e_{31}u_{x,x} + e_{31}u_{y,y} + e_{33}u_{z,z} - \varepsilon_{33}\phi_{,z} - g_{33}\varphi_{,z} \tag{9}$$

$$B_x = d_{15}u_{x,z} + d_{15}u_{z,x} - g_{11}\phi_{,x} - \mu_{11}\varphi_{,x} \tag{10}$$

$$B_y = d_{15}u_{y,z} + d_{15}u_{z,y} - g_{11}\phi_{,y} - \mu_{11}\varphi_{,y} \tag{11}$$

$$B_z = d_{31}u_{x,x} + d_{31}u_{y,y} + d_{33}u_{z,z} - g_{33}\phi_{,z} - \mu_{33}\varphi_{,z} \tag{12}$$

where $u_i$ is the *i-th* component of the elastic displacement vector, and $u_{i,j}$ is the *i,j-th* component of the strain tensor; $\phi$ and $\varphi$ denote the electric and magnetic potentials, respectively; $\sigma_{ij}$, $D_i$ and $B_i$ represent the components of stress tensor, electric displacement vector and magnetic induction vector, respectively; $c_{ij}$, $e_{ij}$ and $d_{ij}$ are the elastic, piezoelectric, and piezomagnetic coefficients, respectively, respectively; $\varepsilon_{ij}$, $g_{ij}$ and $\mu_{ij}$ are the components of the dielectric permittivity, electromagnetic coefficient and magnetic permeability, respectively; Notably, for transversely isotropic materials, the relation of $2c_{66} = c_{11} - c_{12}$ is satisfied.

Since we neglect body forces and electric/magnetic charges, the equilibrium equations and the Maxwell's equations are:

$$\frac{\partial \sigma_{xx}}{\partial x} + \frac{\partial \sigma_{xy}}{\partial y} + \frac{\partial \sigma_{xz}}{\partial z} = 0 \tag{13}$$

$$\frac{\partial \sigma_{yx}}{\partial x} + \frac{\partial \sigma_{yy}}{\partial y} + \frac{\partial \sigma_{yz}}{\partial z} = 0 \tag{14}$$

$$\frac{\partial \sigma_{zx}}{\partial x} + \frac{\partial \sigma_{zy}}{\partial y} + \frac{\partial \sigma_{zz}}{\partial z} = 0 \tag{15}$$

$$\frac{\partial B_x}{\partial x} + \frac{\partial B_y}{\partial y} + \frac{\partial B_z}{\partial z} = 0 \tag{16}$$

$$\frac{\partial D_x}{\partial x} + \frac{\partial D_y}{\partial y} + \frac{\partial D_z}{\partial z} = 0 \tag{17}$$

Referring to **Figures 1**(a) and (b), according to the aforementioned MD adhesive model and Coulomb friction law, the surface contact pressure and shear tractions acting on a generic point (*x,y*) in the calculation domain (*i.e.*, the multiferroic coating surface) are given, respectively, by

$$\begin{cases} p_z(x,y) > -\sigma_0, & g(x,y) = 0 \ \& \ x^2 + y^2 \leq a^2 \\ p_z(x,y) = -\sigma_0, & 0 < g(x,y) \leq h_0 \ \& \ a^2 < x^2 + y^2 \leq c^2 \\ p_z(x,y) = 0, & g(x,y) > h_0 \ \& \ c^2 < x^2 + y^2 \end{cases} \quad (18)$$

and

$$\begin{cases} p_x(x,y) = \mu p_z(x,y), & p_z(x,y) > 0 \\ p_x(x,y) = 0, & p_z(x,y) \leq 0 \end{cases} \quad (19)$$

where

$$g(x,y) = u_z(x,y) + h(x,y) - \delta_z(x,y) \quad (20)$$

where $u_z(x,y)$ is the local gap between the rigid indenter and the deformed coating surface, $u_z(x,y)$ is the coating surface normal displacement, $\delta_z(x,y)$ is the rigid indenter penetration, and $h(x,y)$ is the indenter shape. For the case at hand, we approximate the rigid sphere with an Hertzian indenter; therefore, $h(x,y) = (x^2 + y^2)/2R$, where $R$ is the radius of the sphere. Notably, in the case of multiferroic materials, due to electric-magnetic-elastic coupled behavior, the normal displacement $u_z(x,y)$ not only depends on the distribution of normal pressure $p_z$ and tangential traction $p_x$, but also on the electric charge density $q_b$, and magnetic charge density $g_b$, as further discussed in **Section 2.3**.

The effect of interfacial friction on adhesive interactions is a long-standing, highly debated topic in tribology. Recently, Menga *et al.* [46,47] have rigorously shown that in linear elasticity, for a JKR adhesive contact in gross sliding with uniform frictional stresses at the interface, the overall contact behavior resembles the frictionless case, *i.e.* frictional stresses do not affect adhesion, in agreement with some experimental results [42,48,49]. Nonetheless, a contact area reduction (and shape change) in frictional sliding contact is also reported [50,51] for rubber-like (soft) materials. In these cases, phenomenological models have been suggested, arguing a possible adhesive energy reduction induced by friction, based on empirically tuned arbitrary functions [52,53]; however, accurate FEM simulations [54] and experimental results [55] have recently shown that the observed anisotropic contact shrinking might be related to nonlinear elastic effect (finite deformations) rather than to the adhesion-friction interaction. In this study, since we focus on stiff multiferroic materials in gross slip conditions, we assume no interaction between friction and adhesion, so that the adhesion energy Δγ is constant.

Finally, using **Eqs. (18)-(19)**, the boundary condition of the elastic, electric and magnetic problem at coating surface (*i.e.*, $z=0$) can be expressed as

$$\sigma_{zz}(x,y,0) = -p_z(x,y) \tag{23}$$

$$\sigma_{zx}(x,y,0) = -p_x(x,y) \tag{24}$$

$$\sigma_{zy}(x,y,0) = 0 \tag{25}$$

$$D_z(x,y,0) = \begin{cases} -q_b(x,y), & g(x,y)=0 \\ 0, & g(x,y)>0 \end{cases} \tag{26}$$

$$B_z(x,y,0) = \begin{cases} -g_b(x,y), & g(x,y)=0 \\ 0, & g(x,y)>0 \end{cases} \tag{27}$$

where, since gross slip conditions occur, we neglect the y-component of frictional shear stress, according to the numerical results presented in Ref. [38].

Similarly, at the coating-substrate interface (*i.e.*, $z=h_t$), we have

$$\begin{aligned} u_x\big|_{z=h_t} &= 0, \\ u_y\big|_{z=h_t} &= 0, \\ u_z\big|_{z=h_t} &= 0, \\ \phi\big|_{z=h_t} &= 0, \\ \varphi\big|_{z=h_t} &= 0. \end{aligned} \tag{28}$$

## 2.3 General solutions and Fourier-domain response functions

Substituting the constitutive equations (**Eqs. (1)-(12)**) into the generalized equilibrium equations (**Eqs. (13)-(17)**) results in a set of partial differential equations, with boundary conditions given by **Eqs. (23)-(28)**. The general solutions for the magneto-electro-elastic problem can be derived from this set of equations, following the procedure defined by Ding et al [41-42], which also allows to calculate the corresponding set of Green's functions. Consequently, the displacements field $u$ and the electric $\phi$ and magnetic $\varphi$ potentials fields can be expressed following a Boundary Element Method (BEM) approach as

$$u_x(x,y,z) = \int_{-\infty}^{\infty}\int_{-\infty}^{\infty} \begin{bmatrix} G_{u_x}^{p_x}(x-x',y-y',z)p_x(x',y')+G_{u_x}^{p_z}(x-x',y-y',z)p_z(x',y') \\ +G_{u_x}^{q_b}(x-x',y-y',z)q_b(x',y')+G_{u_x}^{g_b}(x-x',y-y',z)g_b(x',y') \end{bmatrix} dx'dy', \tag{29}$$

$$u_y(x,y,z) = \int_{-\infty}^{\infty}\int_{-\infty}^{\infty} \begin{bmatrix} G_{u_y}^{p_x}(x-x',y-y',z)p_x(x',y')+G_{u_y}^{p_z}(x-x',y-y',z)p_z(x',y') \\ +G_{u_y}^{q_b}(x-x',y-y',z)q_b(x',y')+G_{u_y}^{g_b}(x-x',y-y',z)g_b(x',y') \end{bmatrix} dx'dy', \tag{30}$$

$$u_z(x,y,z) = \int_{-\infty}^{\infty}\int_{-\infty}^{\infty} \begin{bmatrix} G_{u_z}^{p_x}(x-x',y-y',z)p_x(x',y') + G_{u_z}^{p_z}(x-x',y-y',z)p_z(x',y') \\ +G_{u_z}^{q_b}(x-x',y-y',z)q_b(x',y') + G_{u_z}^{g_b}(x-x',y-y',z)g_b(x',y') \end{bmatrix} dx'dy', \quad (31)$$

$$\phi(x,y,z) = \int_{-\infty}^{\infty}\int_{-\infty}^{\infty} \begin{bmatrix} G_{\phi}^{p_x}(x-x',y-y',z)p_x(x',y') + G_{\phi}^{p_z}(x-x',y-y',z)p_z(x',y') \\ +G_{\phi}^{q_b}(x-x',y-y',z)q_b(x',y') + G_{\phi}^{g_b}(x-x',y-y',z)g_b(x',y') \end{bmatrix} dx'dy', \quad (32)$$

$$\varphi(x,y,z) = \int_{-\infty}^{\infty}\int_{-\infty}^{\infty} \begin{bmatrix} G_{\varphi}^{p_x}(x-x',y-y',z)p_x(x',y') + G_{\varphi}^{p_z}(x-x',y-y',z)p_z(x',y') \\ +G_{\varphi}^{q_b}(x-x',y-y',z)q_b(x',y') + G_{\varphi}^{g_b}(x-x',y-y',z)g_b(x',y') \end{bmatrix} dx'dy', \quad (33)$$

where $G_b^a(x,y,z)$ is the Green's function for the quantity $b$ depending on the quantity $a$ (e.g., $G_{u_z}^{q_b}(x,y,z)$ is the Green's function for the normal displacement $u_z$ caused by the surface electric charge $q_b$).

Applying the in-plane Fourier transforms to **Eqs. (29)-(33)** we have

$$\tilde{u}_x(m,n,z) = \begin{bmatrix} \tilde{G}_{u_x}^{p_x} & \tilde{G}_{u_x}^{p_z} & \tilde{G}_{u_x}^{q_b} & \tilde{G}_{u_x}^{g_b} \end{bmatrix} \begin{bmatrix} \tilde{p}_x & \tilde{p}_z & \tilde{q}_b & \tilde{g}_b \end{bmatrix}^T, \quad (34)$$

$$\tilde{u}_y(m,n,z) = \begin{bmatrix} \tilde{G}_{u_y}^{p_x} & \tilde{G}_{u_y}^{p_z} & \tilde{G}_{u_y}^{q_b} & \tilde{G}_{u_y}^{g_b} \end{bmatrix} \begin{bmatrix} \tilde{p}_x & \tilde{p}_z & \tilde{q}_b & \tilde{g}_b \end{bmatrix}^T, \quad (35)$$

$$\tilde{u}_y(m,n,z) = \begin{bmatrix} \tilde{G}_{u_y}^{p_x} & \tilde{G}_{u_y}^{p_z} & \tilde{G}_{u_y}^{q_b} & \tilde{G}_{u_y}^{g_b} \end{bmatrix} \begin{bmatrix} \tilde{p}_x & \tilde{p}_z & \tilde{q}_b & \tilde{g}_b \end{bmatrix}^T, \quad (36)$$

$$\tilde{\phi}(m,n,z) = \begin{bmatrix} \tilde{G}_{\phi}^{p_x} & \tilde{G}_{\phi}^{p_z} & \tilde{G}_{\phi}^{q_b} & \tilde{G}_{\phi}^{g_b} \end{bmatrix} \begin{bmatrix} \tilde{p}_x & \tilde{p}_z & \tilde{q}_b & \tilde{g}_b \end{bmatrix}^T, \quad (37)$$

$$\tilde{\varphi}(m,n,z) = \begin{bmatrix} \tilde{G}_{\varphi}^{p_x} & \tilde{G}_{\varphi}^{p_z} & \tilde{G}_{\varphi}^{q_b} & \tilde{G}_{\varphi}^{g_b} \end{bmatrix} \begin{bmatrix} \tilde{p}_x & \tilde{p}_z & \tilde{q}_b & \tilde{g}_b \end{bmatrix}^T, \quad (38)$$

where hat "$\tilde{\phantom{x}}$" denotes the double Fourier transforms, with $m$ and $n$ for the in-plane wave-numbers. The expressions of the transformed Green's functions of the magneto-electro-elastic systems are given in **Appendix**.

The displacements and electric/magnetic potentials in the spatial domain can be derived in the framework of the discrete convolution-fast Fourier transform (DC-FFT) algorithm [56,57],

$$\begin{bmatrix} u_x & u_y & u_z & \phi & \varphi \end{bmatrix}^T = IFFT\left\{ \begin{bmatrix} \hat{\tilde{C}} \end{bmatrix} \begin{bmatrix} \hat{\tilde{p}}_x & \hat{\tilde{p}}_z & \hat{\tilde{q}}_b & \hat{\tilde{g}}_b \end{bmatrix}^T \right\}, \quad (39)$$

where IFFT is the inverse of the fast Fourier transform (FFT), and

$$[\hat{\hat{C}}] = \begin{bmatrix} \hat{\hat{C}}^{u_x} \\ \hat{\hat{C}}^{u_y} \\ \hat{\hat{C}}^{u_z} \\ \hat{\hat{C}}^{\phi} \\ \hat{\hat{C}}^{\varphi} \end{bmatrix} = \begin{bmatrix} \hat{\hat{C}}^{u_x}_{p_x} & \hat{\hat{C}}^{u_x}_{p_z} & \hat{\hat{C}}^{u_x}_{q_b} & \hat{\hat{C}}^{u_x}_{g_b} \\ \hat{\hat{C}}^{u_y}_{p_x} & \hat{\hat{C}}^{u_y}_{p_z} & \hat{\hat{C}}^{u_y}_{q_b} & \hat{\hat{C}}^{u_y}_{g_b} \\ \hat{\hat{C}}^{u_z}_{p_x} & \hat{\hat{C}}^{u_z}_{p_z} & \hat{\hat{C}}^{u_z}_{q_b} & \hat{\hat{C}}^{u_z}_{g_b} \\ \hat{\hat{C}}^{\phi}_{p_x} & \hat{\hat{C}}^{\phi}_{p_z} & \hat{\hat{C}}^{\phi}_{q_b} & \hat{\hat{C}}^{\phi}_{g_b} \\ \hat{\hat{C}}^{\varphi}_{p_x} & \hat{\hat{C}}^{\varphi}_{p_z} & \hat{\hat{C}}^{\varphi}_{q_b} & \hat{\hat{C}}^{\varphi}_{g_b} \end{bmatrix}, \quad (40)$$

is the matrix of the discrete influence coefficients (ICs) expressed in the Fourier domain, namely the multi-field response of the material under the action of unit point loads, including force, electric, and magnetic loads which can be calculated from the transformed Green's functions [44]. Notably, the hat $\hat{\hat{\phantom{x}}}$ in **Eqs. (39)-(40)** indicates the double FFT with zero padding and ICs wrap around order.

### 2.4 Solution algorithm

According to **Figure 1(a)**, we focus on the sliding contact between a rigid spherical indenter and a multiferroic coating, in the presence of both friction and adhesion. A normal force $P$, and a tangential force $Q$ are applied to the spherical indenter. Surface electric charge density $q_b$, and surface magnetic charge density $g_b$, are applied on the coating surface within the contact regions. Therefore,

$$\int_{S_c} p_z dxdy = P, \int_{S_c} p_x dxdy = Q, \quad (41)$$

$$\int_{S_c} q_b dxdy = Q_b, \int_{S_c} g_b dxdy = G_b, \quad (42)$$

where $p_x$ is given by **Eqs. (19)**, $S_c$ is the whole contact region of radius $c$, $Q_b$ is the surface total electric charge, and $G_b$ is the surface total magnetic charge.

The solution algorithm aims at solving **Eq. (39)** utilizing the DC-FFT numerical scheme [56] for the unknown normal displacement $u_z$, electric potential $\phi$, and magnetic potential $\varphi$. **Eq. (20)** allows to calculate the gap, and iterations can be performed on the normal pressure distribution $p_z(x,y)$ (consistently with **Eq. (18)**) by means of the adhesion-driven conjugate gradient method (AD-CGM) [58,20]. **Figure 2** presents the flow chart for the numerical calculations, with the following steps:

(1) Relevant material properties, contact geometry profiles, convergence accuracy, loading conditions, as well as dimensions of the computational domain, mesh size, friction coefficients, and adhesion parameters, *etc.*, are input.

(2) The ICs for displacements, stresses, electric/magnetic potentials, electric displacements, and magnetic inductions are calculated in advance.

(3) An initial pressure distribution is assumed, and the size of the contact area is specified. The initial surface electric and magnetic charge densities are set to zero.

(4) Surface electric and magnetic charges are applied in incremental steps.

(5) The normal displacement is calculated using the DC-FFT method, and the gap distribution after contact deformation is obtained. The distribution of normal pressure is determined using the AD-CGM algorithm. Force values are checked to ensure they reach the convergence accuracy; if not, the pressure is recomputed.

(6) Steps (4) to (5) are repeated until the surface electric and magnetic charges reach the given values. At this point, the pressure distribution and surface electric/magnetic charge density have been obtained, completing the surface calculations.

(7) Once the $p_x$, $p_z$, $q_b$ and $g_b$ are obtained, the subsurface analysis of the contact problem is carried out utilizing DC-FFT subsequently. Based on the corresponding ICs, the stress components, electric displacements, and magnetic inductions are determined.

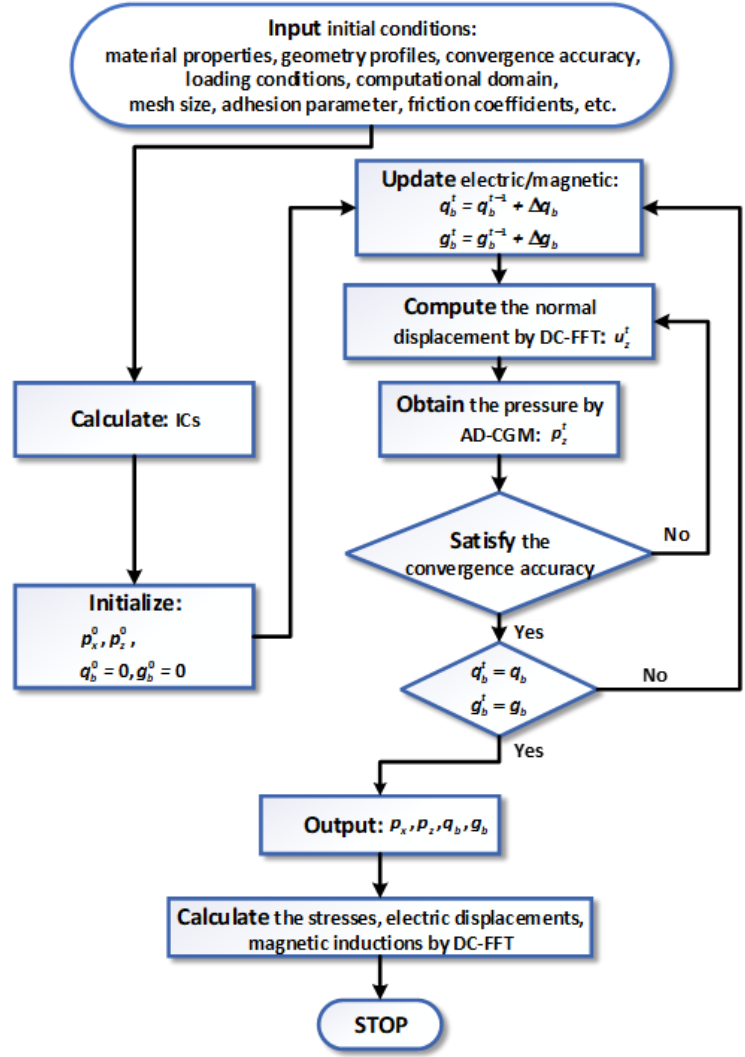

Figure 2 Flow chart for the numerical calculation of the HEM

## 3. Result and discussion

Calculations are performed assuming a multiferroic coating of thickness $h_t$ composed of two phases, a ferromagnetic phase $CoFe_2O_4$ and a ferroelectric phase $BaTiO_3$, forming a transversely isotropic multiferroic material. The chosen material properties in this study possess a distinct set of $s_k$ values (defined in **Appendix**), making the general solutions (**Eq. (29)**) applicable. Specific material properties with a volume fraction 50% of each phase are provided in **Table 1** [59,10], with overall Poisson's ratio $v \approx 0.34$.

**Table 1** The material properties of the transversely isotropic multiferroic coating [59,10] Assume the material properties are independent to the coating thickness. (Units: $c_{ij}$ in $N/m^2$, $e_{ij}$ in $C/m^2$, $d_{ij}$ in $N/Am$, $\varepsilon_{ij}$ in $10^{-9}C^2/Nm^2$, $g_{ij}$ in $10^{-12}Ns/Vc$, $\mu_{ij}$ in $10^{-6}Ns/C^2$)

| $c_{11}$ | $c_{22}$ | $c_{33}$ | $c_{44}$ | $c_{55}$ | $c_{66}$ |
|---|---|---|---|---|---|
| 213.39 | 112.59 | 111.47 | 206.74 | 50.36 | 50.4 |
| $e_{31}$ | $e_{33}$ | $e_{15}$ | $d_{31}$ | $d_{33}$ | $d_{15}$ |
| -2.79 | 8.5 | 0.17 | 224.77 | 290.89 | 163 |
| $\varepsilon_{11}$ | $\varepsilon_{33}$ | $g_{11}$ | $g_{33}$ | $\mu_{11}$ | $\mu_{33}$ |
| 0.23 | 6.38 | 5.53 | 2756.53 | 181 | 88 |

To simplify the reading, we normalize the results with respect to the following reference quantities, calculated for the Hertzian contact of a multiferroic half-space, according to Chen et al. [60] and a MD-typed adhesion model [9,10],

$$a_0 = \sqrt[3]{\frac{3\pi \xi_{11} RP}{4}}, p_0 = \frac{2a_0}{\pi^2 \xi_{11} R},$$

$$\phi_0 = \frac{\xi_{12} a_0^2}{\xi_{11} R}, \varphi_0 = \frac{\xi_{13} a_0^2}{\xi_{11} R}, D_0 = \frac{p_0 e_{33}}{c_{33}}, B_0 = \frac{p_0 d_{33}}{c_{33}},$$

$$\lambda = 2\sigma_0 / \left(\frac{\pi \Delta \gamma H_0^2}{R}\right)^{1/3}, H_0 = 4\eta_{11}/(3\pi\eta) \tag{45}$$

where $a_0$ is the reference contact radius, $p_0$ is the reference maximum contact pressure, $\phi_0$ is the reference equivalent electric potential, $\varphi_0$ is the reference equivalent magnetic potential, $D_0$ is the reference equivalent electric displacement, and $B_0$ is the reference equivalent magnetic induction. Similarly, $\lambda$ is the Maugis adhesion parameter [9]. Moreover, in Eq. (45), the parameters $\xi_{11} = 2.03552\times 10^{-12} \text{m}^2/\text{N}$, $\xi_{12} = 1.77480 \times 10^{-13} \text{m}^2/\text{C}$, $\xi_{13} = 1.93900\times 10^{-6}\text{mA/N}$, $\eta = 5.14794\times 10^{-2}\text{m}^4\text{A}^2/(\text{NC}^2)$ and $\eta_{11} = 2.36834\times 10^{10}\text{m}^2\text{A}^2/\text{C}^2$ are given in Chen et al. [60] based on the material properties. Notably, $H_0$ in Eq. (45) correspond to in the stiffness parameter $K$ in Refs. [9,10]. In what follows, we set the computational domain size in $x$, $y$, and $z$ directions as $3a_0 \times 3a_0 \times h_t$ with a grid resolution of $256\times 256\times 64$. Moreover, if not explicitly indicated, the indenter radius is $R = 10\text{mm}$, and the surface electric and magnetic charge density are $q_b = 0.001\text{C/m}^2$ and $g_b = 0.001\text{N/Am}$, respectively.

## 3.1 Model validation

Model validation against analytical solutions by Wu *et al.* [10] can be performed in the case of adhesive frictionless contact of an insulating Hertzian indented and a multiferroic half-space. Consequently, we let the friction coefficient vanish (*i.e.*, $\mu=0$) and the coating thickness tend to infinite (*i.e.*, $h_t \square R \square a_0$). Moreover, we set the adhesion parameter $\lambda=0.5$, and different values of the applied normal load, *i.e.*, $P=0.5\text{N}$, $1.0\text{N}$ and $2.0\text{N}$. The dimensionless normal displacement $10^5 u_z/R$ comparison shown in **Figure 3** indicates that the numerical model is in very good agreement with theoretical predictions [10], regardless of the value of *P*.

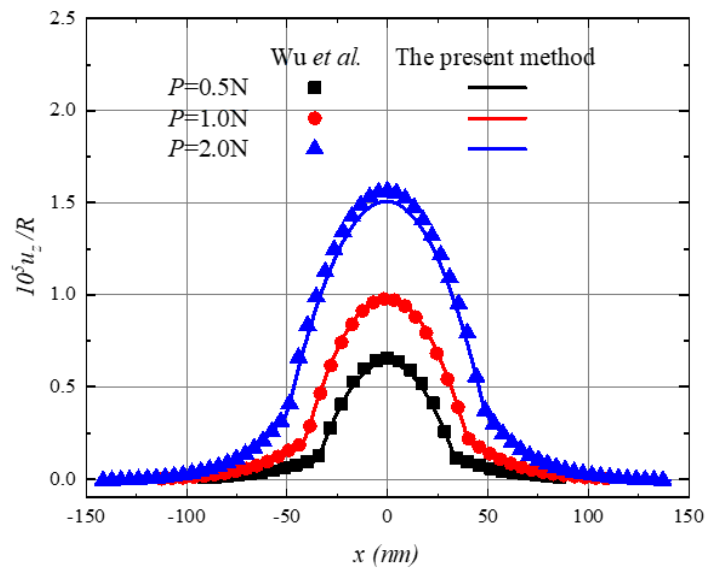

**Figure 3** Validation of the present method by comparison with the solutions by Wu *et al.* [9] for the problem of a multiferroic half-space indented by a rigid insulating sphere. The indenter radius is $R=10\text{mm}$, and the external normal loads *P* varies among 0.5, 1.0 and 1.5N.

### 3.2 Effects of adhesion parameters

We expect the adhesion parameter $\lambda$ to significantly affects the contact behavior. To investigate this case, we focus on different values of the adhesion parameter $\lambda$. The normal load acting on the indenter is $P=100\text{mN}$, and the friction coefficient $\mu=0.6$. We set the multiferroic coating thickness $h_t=1.0 a_0$, in-plane/out-of-plane elastic coupling [36–38] is very little and mostly related to compressibility effects ($v \approx 0.34$), while geometric (*i.e.*, confinement) coupling is negligible.

The adhesion parameter $\lambda$ affects the pressure distribution according to MD adhesion model. Specifically, as shown in **Figure 4**(a), increasing $\lambda$ leads to thinner annular adhesive region and higher adhesive stresses, and vice versa. In the limit of $\lambda \gg 1$, the JKR [7] behavior is asymptotically approached (*i.e.*, singular adhesive stresses); while, for $\lambda \ll 1$, the DMT [8] behavior is recovered. From **Figures 4**(b) and (c), it can be observed that surface electric and magnetic potential distributions of the coating surfaces become asymmetrical due to interfacial friction. When adhesive forces are considered, the electric and magnetic potentials within the adhesive regions are decreased compared to the adhesiveless case, and a pronounced sharp downward protrusion is observed when $\lambda > 0.5$. Overall, as the value of $\lambda$ approaches zero, electric and magnetic potentials exhibit a smoother behavior, though still asymmetric.

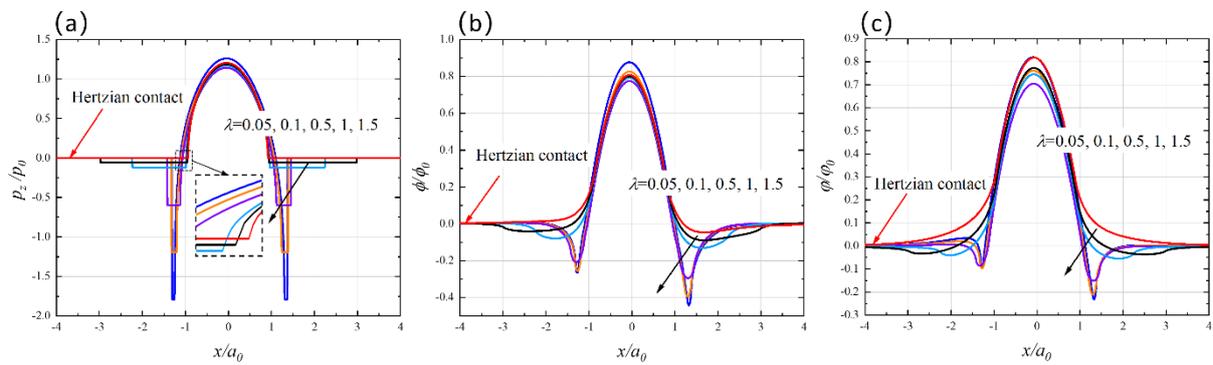

**Figure 4** Cross-section of the (a) dimensionless contact pressure distribution $p_z / p_0$, (b) dimensionless contact electric potential distribution $\phi / \phi_0$, (c) dimensionless contact magnetic potential distribution $\varphi / \varphi_0$, for different values of $\lambda$. The indenter is sliding from left to right. Calculations refer to $R = 10$ mm, $h_t = 1.0 a_0$, $P = 100$ mN, $\mu = 0.6$; therefore, $a_0 = 1.686 \times 10^{-5}$ m, $p_0 = 1.679 \times 10^8$ Pa, $\phi_0 = 72.507$ V, and $\varphi_0 = 2.709 \times 10^{-2}$ A.

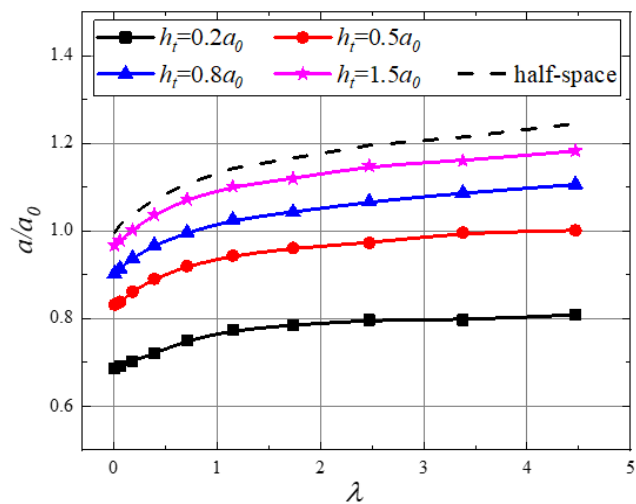

**Figure 5** Contact radius as a function of $\lambda$, for different coating thickness. Frictionless conditions are assumed. The normal load is $P = 100$ mN.

The effect of the adhesion parameter $\lambda$ on the dimensionless contact radius $a/a_0$ for various coating thickness $h_t$ is plotted in **Figure 5**, under frictionless conditions. As expected, increasing $\lambda$ leads to larger contact area. Notably, for the value of $P$ under investigation, very low values of $\lambda$ correspond to the (Hertzian) adhesiveless behavior $a/a_0 \approx 1$, while lower normal loads would lead to $a/a_0 > 1$ even for $\lambda \ll 1$.

The influence of the adhesive parameter $\lambda$ on subsurface quantities is investigated for both frictionless ($\mu=0.0$) and frictional ($\mu=0.3$) conditions. Specifically, **Figures 6-8**, show the normalized distributions of stress $\sigma_{zz}/p_0$, electric displacement $D_z/D_0$, and magnetic induction $B_z/B_0$ in the z-direction of the subsurface x-z cross-section, respectively. Again, the effect of $\lambda$ is symmetric, with the contact area and general fields features enhanced by adhesion. More importantly, friction breaks the symmetry of all the investigated fields [38], leading to a significant concentration of $D_z/D_0$ and $B_z/B_0$ beneath the contact trailing edge. Moreover, the minimum electric displacement is observed at the bottom of the coating in the contact area, and the minimum magnetic induction is found near the surface close to the contact area.

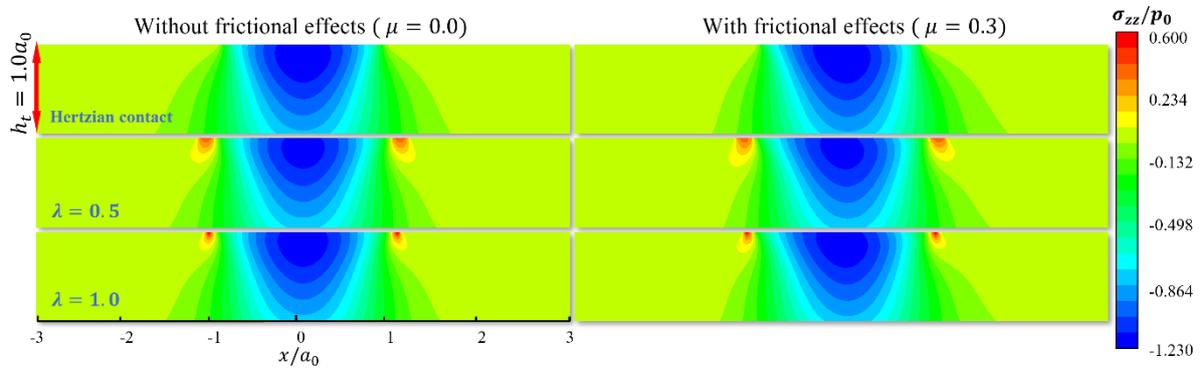

**Figure 6** Subsurface contour map of dimensionless z-directional stress $\sigma_{zz}/p_0$ on the x-z section under different adhesion parameters and friction coefficients. The indenter is sliding from left to right. In this case, the indenter radius is $R=10$mm, the coating thickness is $h_t=1.0a_0$, the external normal load is $P=500$mN, the referenced values are $a_0=2.884\times10^{-5}$m, $p_0=2.871\times10^8$Pa, $\phi_0=72.507$V, $\varphi_0=7.921\times10^{-2}$A, $D_0=1.180\times10^{-2}$C/m$^2$, and $B_0$=0.404N/Am.

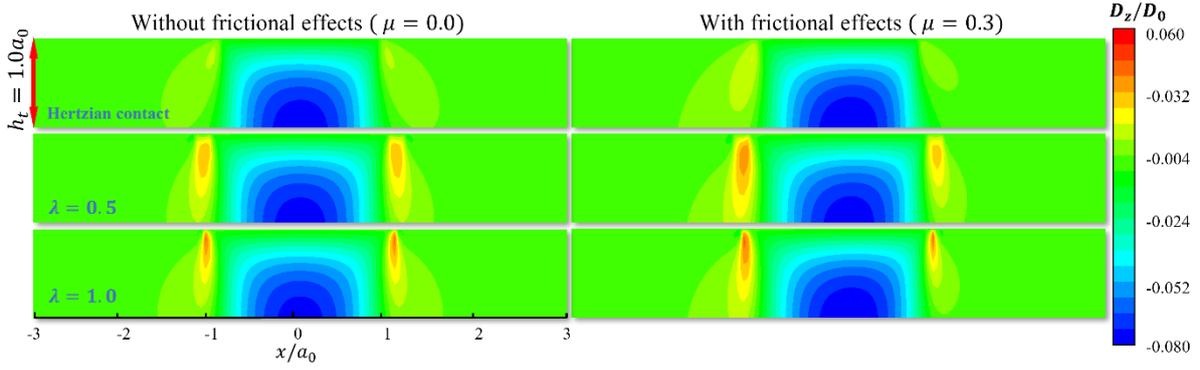

**Figure 7** Subsurface contour map of dimensionless z-direction electric displacement $D_z/D_0$ on the x-z section under different adhesion parameters and friction coefficients. The indenter is sliding from left to right.

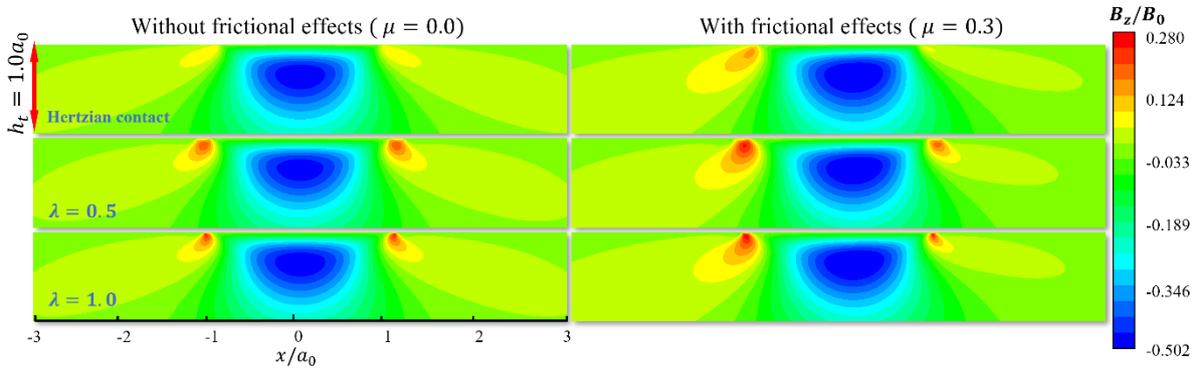

**Figure 8** Subsurface contour map of dimensionless z-directional magnetic induction $B_z/B_0$ on the x-z section under different adhesion parameters and friction coefficients. The indenter is sliding from left to right.

### 3.3 Effects of coating thickness

Reducing the multiferroic coating thickness increases the contact stiffness [61] and the degree of in-plane/out-of-plane geometric coupling [36–38]. In this section, we consider a coating thickness $h_t$ varying from $0.2a_0$ to $1.5a_0$. The normal load is set to $P=500\text{mN}$, with $\lambda=0.5$, and $\mu=0.6$.

**Figure 9** confirms the strong effect of the coating thickness on the contact pressure $p_z$, surface electric potential $\phi$, and surface magnetic potential $\varphi$. Specifically, in agreement with [61], given the normal contact force acting on the indenter, reducing the coating thickness $h_t$ increases the peak pressure and reduces the contact area radius $a$. A deeper analysis of **Figure 9**(a) reveals that the coating thickness also affect the friction-induced pressure asymmetry, as the pressure peak shifts towards the leading edge for thinner coatings. Moreover, as discussed in **Section 2.2**, frictional coupling also affects gap distribution and, in turn, adhesion.

For very thick layers, **Figure 9**(a) shows that the adhesive region is larger at the trailing edge due to contact asymmetry induced by compressibility (*i.e.*, material) coupling; as expected, this effect reduces with $h_t$ reducing, as geometric coupling increases balancing the material one (see Refs. [36–38] for further details).

Overall, the electric potential $\phi$ and the magnetic potential $\varphi$ increase with the coating thickness throughout the entire computational domain.

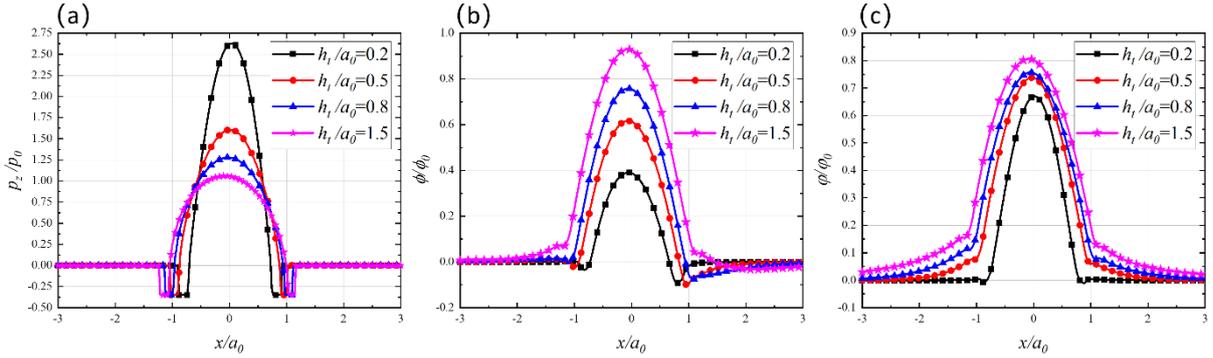

**Figure 9** Cross-section for different values of coating thickness $h_t$ of the surface value of: (a) dimensionless contact pressure distribution $p_z/p_0$; (b) dimensionless contact electric potential distribution $\phi/\phi_0$; (c) dimensionless contact magnetic potential distribution $\varphi/\varphi_0$. In this case, the indenter radius is $R=10\text{mm}$, the external normal load is $P=500\text{mN}$, the adhesion parameter is $\lambda=0.5$, and the friction coefficient is $\mu=0.6$. The referenced parameters are $a_0=2.884\times10^{-5}\text{m}$, $p_0=2.871\times10^8\text{Pa}$, $\phi_0=72.507\text{V}$, and $\varphi_0=7.921\times10^{-2}\text{A}$.

### 3.4 Effects of friction coefficient

In this section, we explore the effect of the friction coefficient value (namely, we set $\mu=0.0, 0.3, 0.6, 0.9$), assuming the applied normal load $P=500\text{mN}$ and the adhesion parameter of $\lambda=0.5$. Calculations refer to a coating thickness $h_t=0.2a_0$. **Figure 10** illustrates the distributions of surface pressure $p_z$, electric potential $\phi$, and magnetic potential $\varphi$ under different friction coefficients. As expected, with an adhesion coefficient of $\lambda=0.5$, an adhesive annular region forms outside the contact area. Similarly to what observed in **Figure 9**(a), increasing the friction coefficient leads stronger geometric in-plane/out-of-plane coupling, which shifts the pressure distribution towards the contact leading edge. All the same, the leading and trailing edges values of the electric and magnetic potentials consistently increase and decrease, respectively, with the friction coefficient.

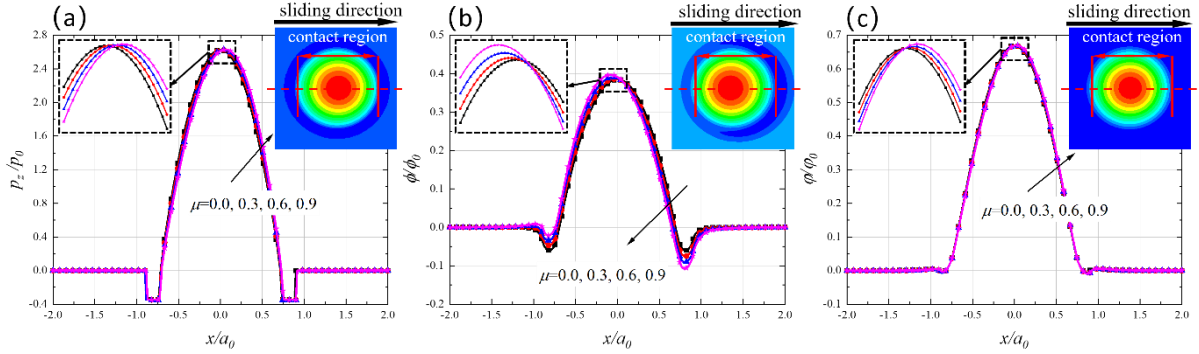

**Figure 10** Normalized surface contact behaviors for different values of friction coefficient $\mu$: (a) dimensionless contact pressure distribution $p_z/p_0$ ($p_0$ is the referenced maximum contact pressure, $a_0$ is the referenced contact radius); (b) dimensionless contact electric potential distribution $\phi/\phi_0$ ($\phi_0 = \xi_{12} a_0^2 / \xi_{11} R$); (c) dimensionless contact magnetic potential distribution $\varphi/\varphi_0$ ($\varphi_0 = \xi_{13} a_0^2 / \xi_{11} R$). In this case, the indenter radius is $R=10$mm, the coating thickness is $h_t = 1.0 a_0$, the external normal load is $P=500$mN, the referenced parameters are $a_0 = 2.884 \times 10^{-5}$m, $p_0 = 2.871 \times 10^8$Pa, $\phi_0 = 72.507$V, $\varphi_0 = 7.921 \times 10^{-2}$A.

### 3.5 Effects of surface electric and magnetic charges

The effect of the surface electric $q_b$ and magnetic $g_b$ charges is investigated assuming $P=500$mN, $h_t = 1.0 a_0$, and $\lambda = 0.5$. Also, we set the friction coefficient $\mu = 0.6$. The contact pressure distribution $p_z$ is mostly unaffected. Indeed, **Figures 11**(a) only shows a slight reduction of the contact peak pressure with both $q_b$ and $g_b$ increasing. On the contrary, the surface electric $\phi$ and magnetic $\varphi$ potentials are strongly affected. More in details, **Figures 11**(b) and (c) show that, in general, $\phi$ and $\varphi$ do not vanish at the contact edges and are finite even in the non-contact region. Moreover, in the range of values here investigated, increasing the surface electric/magnetic charges, the electric potential significantly diminishes, and the magnetic potential slightly increases.

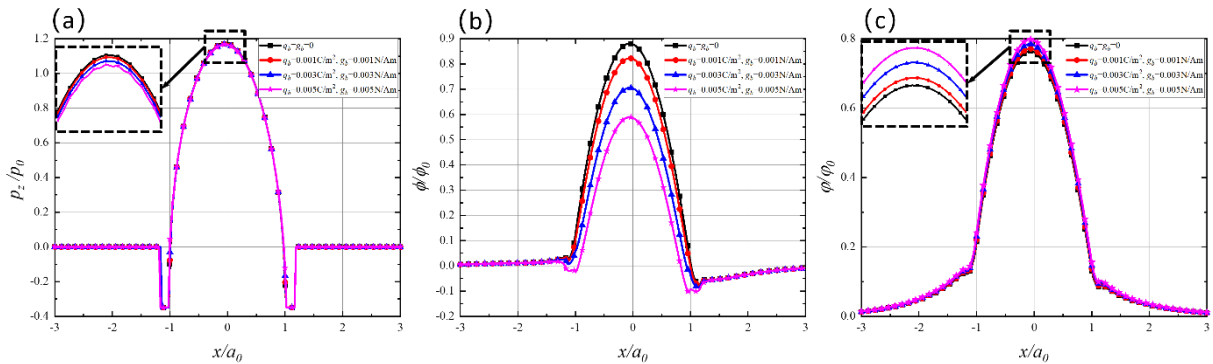

**Figure 11** Normalized surface contact behaviors for different values of electric charge density $q_b$ and magnetic charge density $g_b$: (a) dimensionless contact pressure distribution $p_z / p_0$ ($p_0$ is the referenced maximum contact pressure, $a_0$ is the referenced contact radius); (b) dimensionless contact electric potential distribution $\phi / \phi_0$ ($\phi_0 = \xi_{12} a_0^2 / \xi_{11} R$); (c) dimensionless contact magnetic potential distribution $\varphi / \varphi_0$ ($\varphi_0 = \xi_{13} a_0^2 / \xi_{11} R$). In this case, the indenter radius is $R = 10\text{mm}$, the coating thickness is $h_t = 1.0 a_0$, the external normal load is $P = 500\text{mN}$, the referenced parameters are $a_0 = 2.884 \times 10^{-5}\text{m}$, $p_0 = 2.871 \times 10^8 \text{Pa}$, $\phi_0 = 72.507\text{V}$, and $\varphi_0 = 7.921 \times 10^{-2} \text{A}$.

## 3.6 Subsurface von Mises stress

Plasticity is governed by the von Mises stress distribution,

$$\sigma_{VM} = \sqrt{[(\sigma_{xx} - \sigma_{yy})^2 + (\sigma_{yy} - \sigma_{zz})^2 + (\sigma_{zz} - \sigma_{xx})^2 + 6(\sigma_{xy}^2 + \sigma_{xz}^2 + \sigma_{yz}^2)]/2} \qquad (46)$$

Therefore, the analysis of $\sigma_{VM}$ may help in preventing plastic deformation spots in multiferroic coatings, which might eventually affect fatigue and material failure (*e.g.* crack nucleation and propagation).

**Figure 12** shows the combined effects of adhesion and friction on subsurface von Mises stress in the *x-z* cross-section of the multiferroic coatings. In the frictionless case, the von Mises stress field is always symmetric, with the maximum located in the bulk of the coating, beneath the contact surface, where the material is highly compressed; in this case (*i.e.* $\mu = 0$), the effect of cohesive stresses for $\lambda = 0.5$ is to increase the $\sigma_{VM}$ beneath the contact edges. Regardless of the adhesive behavior, introducing frictional shear stresses in the contact zone leads to an asymmetric von Mises stress field: the higher the friction coefficient, the higher the degree of stress field asymmetry. More in detail, increasing the value of $\mu$ shifts the maximum value of $\sigma_{VM}$ closer to the contact trailing edge and to the coating surface. In this regard, the presence of adhesion enhances this behavior.

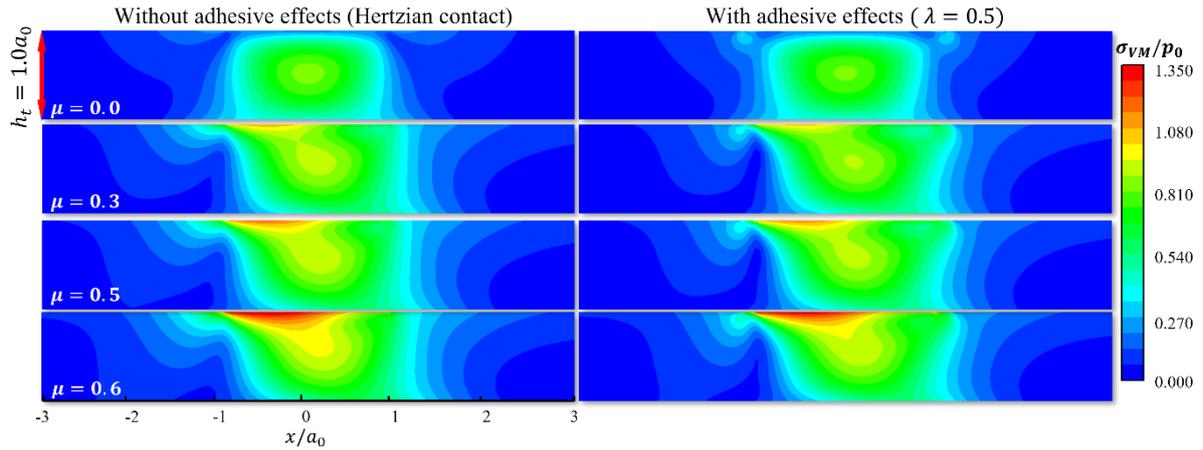

**Figure 12** Subsurface contours of the dimensionless von Mises stresses $\sigma_{VM}/p_0$ in the x-z cross-section for different values friction coefficient $\mu$ and with/without adhesion effects. The indenter is sliding from left to right. Calculations refer to $R=10$mm, $h_t=1.0a_0$, $P=500$mN; therefore, $a_0=2.884\times 10^{-5}$m, $p_0=2.871\times 10^8$Pa, $\phi_0=72.507$V, and $\varphi_0=7.921\times 10^{-2}$A.

The effect of adhesion is even more clearly reported in **Figure 13**, showing that for very large value of $\lambda$ (*i.e.* for very high cohesive stress $\sigma_0$), von Mises stress concentration also occurs in a small spot at the leading edge (see **Figure 13**(b)). On the contrary, the value of $\sigma_{VM}$ at the contact area surface is poorly affected by adhesion.

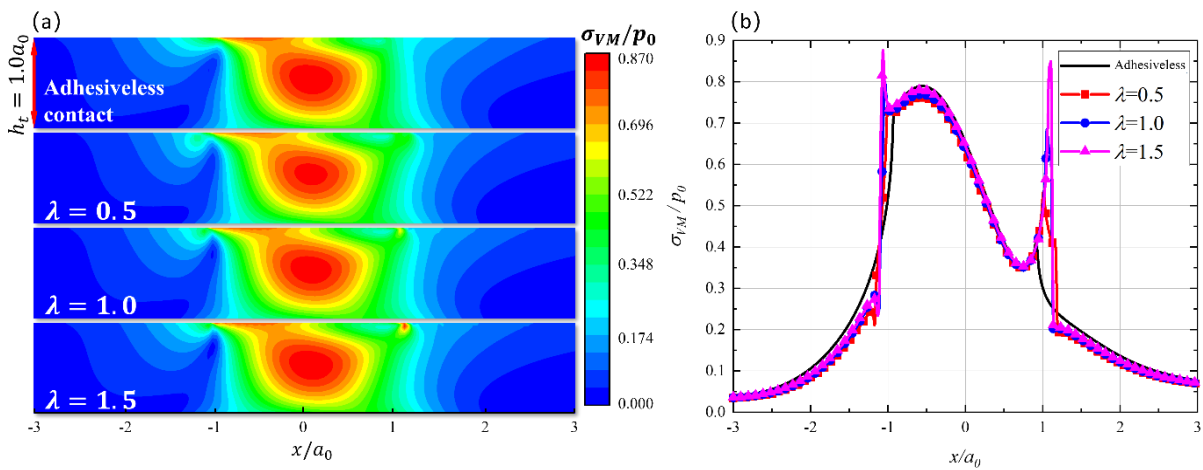

**Figure 13** (a) Subsurface contours of the dimensionless von Mises stresses $\sigma_{VM}/p_0$ in the x-z cross-section and (b) surface distribution of $\sigma_{VM}/p_0$ for different values of adhesion parameter $\lambda$. The indenter is sliding from left to right. Calculations refer to $R=10$mm, $h_t=1.0a_0$, $P=500$mN, $\mu=0.3$; therefore, $a_0=2.884\times 10^{-5}$m, $p_0=2.871\times 10^8$Pa, $\phi_0=72.507$V, and $\varphi_0=7.921\times 10^{-2}$A.

## 3.7 Effects of normal load

The effect of the normal load on the contact behavior is shown in **Figure 14** for both adhesiveless and different adhesive conditions. In the presence of adhesion, the value of the adhesive parameter $\lambda$ controls the intensity of the cohesive stress acting in the annular region surrounding the contact. As expected, the overall effect of adhesion is to increase the contact radius $a$ at a given applied normal force; moreover, increasing $\lambda$ also increases the value of $a$. Furthermore, increasing the applied normal load $P$, the size of the cohesive region (*i.e.* $c-a$) decreases. As for **Figure 4**(a), we also observe a slight shift of the peak pressure towards the trailing edge due to the presence of compressibility-induced in-plane/out-of-plane coupling [36–38], therefore the adhesiveless solution here does not formally correspond to the Hertzian one.

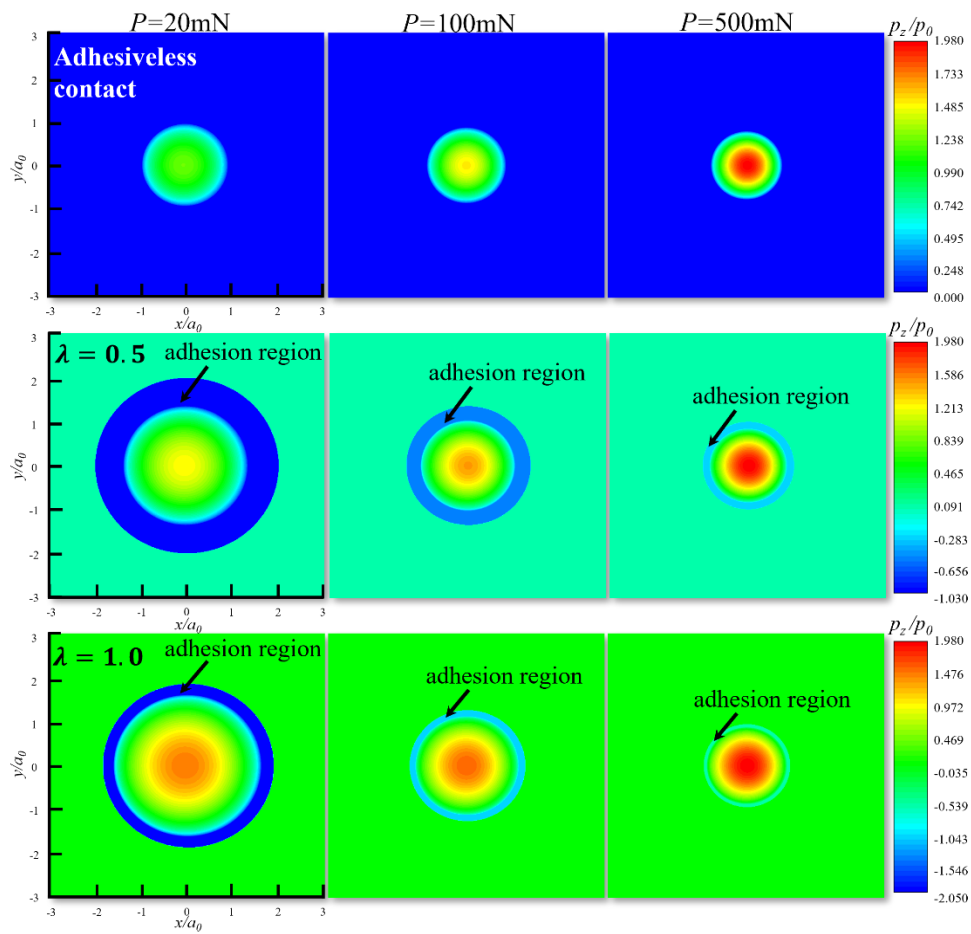

**Figure 14** Dimensionless contact pressure distribution $p_z/p_0$ for different values of adhesion parameter $\lambda$ and external normal load $P$. The indenter is sliding from left to right. Calculations refer to $R=10$mm, $h_t=10$μm, and $\mu=0.3$. Reference values are: $a_0=0.986\times10^{-5}$m and $p_0=0.982\times10^8$Pa for $P=50$mN ; $a_0=1.686\times10^{-5}$m and $p_0=1.679\times10^8$Pa for $P=100$mN ; $a_0=2.884\times10^{-5}$m and $p_0=2.871\times10^8$Pa for $P=500$mN.

To better investigate the effect of the applied normal load and adhesive conditions, we define the adhesive ratio $m=c/a$, where $c$ and $a$ are the outer and inner radii of the adhesive annular region, respectively. A smaller value of $m$ indicates that the adhesive region occupies a smaller proportion of the total contact area, $S_c$ (defined in **Eq. 42**). **Figure 15** depicts the adhesive ratio $m=c/a$ as a function of the applied normal load $P$ under different values of the adhesion parameter and coating thickness. To measure the symmetric contact region, friction is set to zero here. As observed in **Figure 14**, increasing $P$ leads to a reduction of the adhesive ratio $m=c/a$, in agreement with Maugis [9] and Wu *et al.* [10] predictions for semi-infinite elastic and multiferroic solids, respectively. However, **Figure 15**(b) also reveals that thinner coatings present higher adhesive ratios. This interesting result is related to different features of confined solids of finite thickness: (i) the overall stiffer contact behavior resulting is a smaller contact radius $a$; and (ii) the different gap distribution close to the contact edges [36], compared to the case of semi-infinite solids.

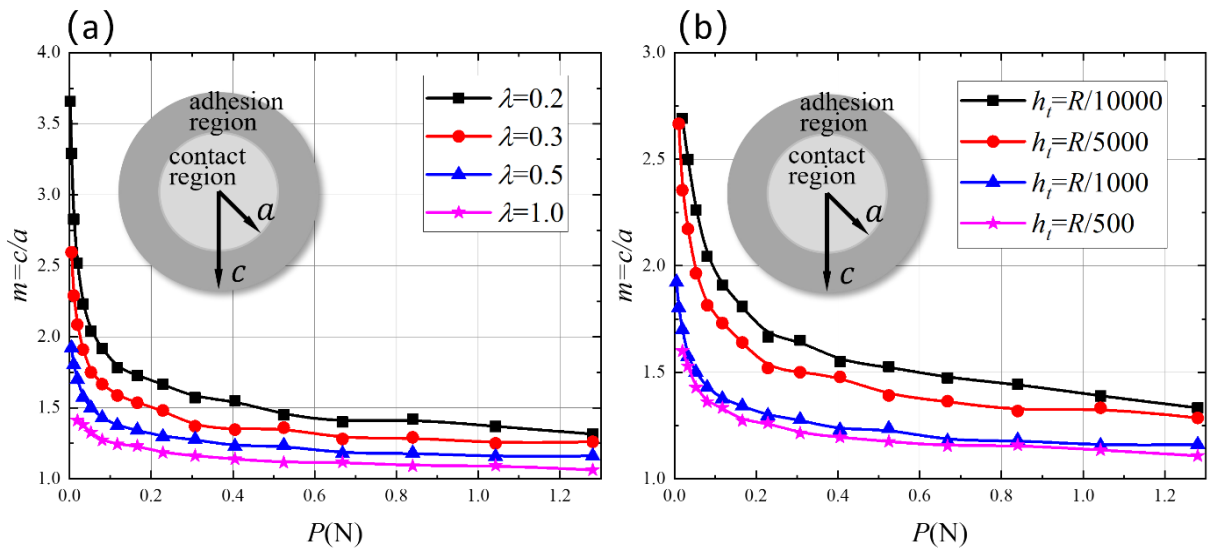

**Figure 15** Relation between the adhesive ratio $m=c/a$ and the external load $P$ for various adhesion parameter $\lambda$ and coating thickness $h_t$. In this case, the indenter radius keeps a constant of $R=10$mm. Calculations here refer to frictionless contacts.

## 4. Conclusions

We studied the frictional adhesive contact of a rigid insulating Hertzian indenter sliding past a multiferroic coating deposed onto a rigid substrate. The problem formulation relies on the hybrid element method (HEM), so that the core analytical solutions for the magneto-electro-elastic coating are expressed in the Fourier domain (in x and y directions), and the condensed into influence coefficients (ICs). Firstly, the unknown contact pressure distribution is found using the adhesion-driven conjugate gradient method (AD-CGM), then the normal displacements, surface electric and magnetic potentials, subsurface stresses, electric displacements, and subsurface magnetic inductions are calculated using the discrete convolution-fast Fourier transform (DC-FFT) algorithm.

We focus our analysis on the effects of the adhesion parameter $\lambda$, the friction coefficient $\mu$, the coating thickness $h_t$, the surface charge density $q_b$, the surface magnetic density $g_b$, and the indenter normal force $P$ on the overall contact behavior. Numerical results allow to draw the following conclusions:

(1) Due to MD adhesion, the surface pressure distribution presents an annular cohesive region surrounding the (repulsive) contact region. As the adhesion parameter $\lambda$ increases, the radial size of the cohesive region reduces, while the adhesive tractions increase. Increasing the normal force $P$ on the indenter also reduces the adhesive region size. Increasing the coating thickness $h_t$ leads to a larger contact area, given the same applied normal force, while the effect of $h_t$ on the adhesive region is poor. Furthermore, increasing the surface electric and/or magnetic charge densities $q_b$ and $g_b$ contribute to a decrease in the surface pressure.

(2) Adhesion and friction primarily exert an influence on the electric and magnetic behavior close to the adhesion region. Specifically, the presence of adhesion leads to a reduction in the electric and magnetic potentials in the annular cohesive region outside the contact area. Adhesion also increases the subsurface (z direction), electric displacement and magnetic induction. For thin coating, friction breaks the symmetry in all the fields, increasing (decreasing) the absolute value of electric and magnetic potentials at the leading (trailing) edge. Regarding the subsurface behavior, friction weakens the electric displacement and magnetic induction at the leading edge, while enhancing those at the trailing edge. Greater surface charge density $q_b$ and magnetic charge density $g_b$ predominantly affect a reduction in electric potential and an increase in magnetic potential within the contact area.

(3) Adhesion increases the subsurface stress (in z direction) and the von Mises stress in the proximity of the adhesion region; however, this effect reduces moving deeper in the coating. For sufficiently thin multiferroic coatings, friction leads to a non-circular contact area and induces asymmetry in the surface displacements, contact pressure, and subsurface stresses (both in z direction and von Mises stress). Interestingly, in agreement with Ref [39], friction leads to a stress concentration at the trailing edge of the contact.

## Acknowledgments

Part pf the authors (TL, BP, YT, LM, XZ) would like to acknowledge support by the National Natural Science Foundation of China (12102085), and the Postdoctoral Science Foundation of China (2023M730504). This work was partly supported by the European Union – NextGenerationEU through the Italian Ministry of University and Research under the programs: (NM) PRIN2022 (Projects of Relevant National Interest) grant nr. 2022SJ8HTC - ELectroactive gripper For mIcro-object maNipulation (ELFIN); (NM) PRIN2022 PNRR (Projects of Relevant National Interest) grant nr. P2022MAZHX - TRibological modellIng for sustainaBle design Of induStrial friCtiOnal inteRfacEs (TRIBOSCORE).

## Appendix

The Fourier transformed Green's functions represent the multi-field response to multi-field loadings, including force, electric, and magnetic loads. Therefore, by setting $p_z = 1$, $q_b = 1$, $g_b = 1$ in the multi-field responses given by **Eqs. (34-38)**, a series of Fourier transformed Green's functions for the displacements $u_i$ and electric/magnetic potentials, $\phi$ and $\varphi$, respectively, as shown in **Eq. (A1)**

$$[\tilde{\tilde{G}}] = \begin{bmatrix} \tilde{\tilde{G}}^{u_x} \\ \tilde{\tilde{G}}^{u_y} \\ \tilde{\tilde{G}}^{u_z} \\ \tilde{\tilde{G}}^{\phi} \\ \tilde{\tilde{G}}^{\varphi} \end{bmatrix} = \begin{bmatrix} \tilde{\tilde{G}}^{u_x}_{\tilde{\tilde{p}}_x=\mu} & \tilde{\tilde{G}}^{u_x}_{\tilde{\tilde{p}}_z=1} & \tilde{\tilde{G}}^{u_x}_{\tilde{\tilde{q}}_b=1} & \tilde{\tilde{G}}^{u_x}_{\tilde{\tilde{g}}_b=1} \\ \tilde{\tilde{G}}^{u_y}_{\tilde{\tilde{p}}_x=\mu} & \tilde{\tilde{G}}^{u_y}_{\tilde{\tilde{p}}_z=1} & \tilde{\tilde{G}}^{u_y}_{\tilde{\tilde{q}}_b=1} & \tilde{\tilde{G}}^{u_y}_{\tilde{\tilde{g}}_b=1} \\ \tilde{\tilde{G}}^{u_z}_{\tilde{\tilde{p}}_x=\mu} & \tilde{\tilde{G}}^{u_z}_{\tilde{\tilde{p}}_z=1} & \tilde{\tilde{G}}^{u_z}_{\tilde{\tilde{q}}_b=1} & \tilde{\tilde{G}}^{u_z}_{\tilde{\tilde{g}}_b=1} \\ \tilde{\tilde{G}}^{\phi}_{\tilde{\tilde{p}}_x=\mu} & \tilde{\tilde{G}}^{\phi}_{\tilde{\tilde{p}}_z=1} & \tilde{\tilde{G}}^{\phi}_{\tilde{\tilde{q}}_b=1} & \tilde{\tilde{G}}^{\phi}_{\tilde{\tilde{g}}_b=1} \\ \tilde{\tilde{G}}^{\varphi}_{\tilde{\tilde{p}}_x=\mu} & \tilde{\tilde{G}}^{\varphi}_{\tilde{\tilde{p}}_z=1} & \tilde{\tilde{G}}^{\varphi}_{\tilde{\tilde{q}}_b=1} & \tilde{\tilde{G}}^{\varphi}_{\tilde{\tilde{g}}_b=1} \end{bmatrix} \qquad (A1)$$

The Fourier transforms of displacements, electric potentials, magnetic potentials, stresses, electric displacements, and magnetic inductions are expressed as:

$$\tilde{\tilde{u}}_x = -inA_0 e^{-\alpha s_0 z} - in\bar{A}_0 e^{\alpha s_0 z} + \sum_{j=1}^{4} imA_j e^{-\alpha s_j z} + \sum_{j=1}^{4} im\bar{A}_j e^{\alpha s_j z},$$

$$\tilde{\tilde{u}}_y = imA_0 e^{-\alpha s_0 z} + im\bar{A}_0 e^{\alpha s_0 z} + \sum_{j=1}^{4} inA_j e^{-\alpha s_j z} + \sum_{j=1}^{4} in\bar{A}_j e^{\alpha s_j z},$$

$$\tilde{\tilde{u}}_z = -\sum_{j=1}^{4} s_j k_{1j} \alpha A_j e^{-\alpha s_j z} + \sum_{j=1}^{4} s_j k_{1j} \alpha \bar{A}_j e^{\alpha s_j z}, \tag{A2}$$

$$\phi = -\sum_{j=1}^{4} s_j k_{2j} \alpha A_j e^{-\alpha s_j z} + \sum_{j=1}^{4} s_j k_{2j} \alpha \bar{A}_j e^{\alpha s_j z},$$

$$\tilde{\tilde{\varphi}} = -\sum_{j=1}^{4} s_j k_{3j} \alpha A_j e^{-\alpha s_j z} + \sum_{j=1}^{4} s_j k_{3j} \alpha \bar{A}_j e^{\alpha s_j z},$$

$$\tilde{\tilde{\sigma}}_{xx} = 2c_{66} mn A_0 e^{-\alpha s_0 z} + \sum_{j=1}^{4} [(c_{66} - s_j^2 \omega_{1j})\alpha^2 + c_{66}(n^2 - m^2)] A_j e^{-\alpha s_j z}$$

$$+ 2c_{66} mn \bar{A}_0 e^{\alpha s_0 z} + \sum_{j=1}^{4} [(c_{66} - s_j^2 \omega_{1j})\alpha^2 + c_{66}(n^2 - m^2)] \bar{A}_j e^{\alpha s_j z},$$

$$\tilde{\tilde{\sigma}}_{yy} = -2c_{66} mn A_0 e^{-\alpha s_0 z} + \sum_{j=1}^{4} [(c_{66} - s_j^2 \omega_{1j})\alpha^2 - c_{66}(n^2 - m^2)] A_j e^{-\alpha s_j z}$$

$$- 2c_{66} mn \bar{A}_0 e^{\alpha s_0 z} + \sum_{j=1}^{4} [(c_{66} - s_j^2 \omega_{1j})\alpha^2 - c_{66}(n^2 - m^2)] \bar{A}_j e^{\alpha s_j z},$$

$$\tilde{\tilde{\sigma}}_{zz} = \sum_{j=1}^{4} \omega_{1j} \alpha^2 A_j e^{-\alpha s_j z} + \sum_{j=1}^{4} \omega_{1j} \alpha^2 \bar{A}_j e^{\alpha s_j z},$$

$$\tilde{\tilde{\sigma}}_{xy} = c_{66}(n^2 - m^2) A_0 e^{-\alpha s_0 z} - \sum_{j=1}^{4} 2c_{66} mn A_j e^{-\alpha s_j z}$$

$$+ c_{66}(n^2 - m^2) \bar{A}_0 e^{\alpha s_0 z} - \sum_{j=1}^{4} 2c_{66} mn \bar{A}_j e^{\alpha s_j z},$$

$$\tilde{\tilde{\sigma}}_{zx} = ins_0 \rho_1 \alpha A_0 e^{-\alpha s_0 z} - \sum_{j=1}^{4} ims_j \omega_{1j} \alpha A_j e^{-\alpha s_j z}$$

$$- ins_0 \rho_1 \alpha \bar{A}_0 e^{\alpha s_0 z} + \sum_{j=1}^{4} ims_j \omega_{1j} \alpha \bar{A}_j e^{\alpha s_j z}, \tag{A3}$$

$$\tilde{\tilde{\sigma}}_{zy} = -ims_0 \rho_1 \alpha A_0 e^{-\alpha s_0 z} - \sum_{j=1}^{4} ins_j \omega_{1j} \alpha A_j e^{-\alpha s_j z}$$

$$+ ims_0 \rho_1 \alpha \bar{A}_0 e^{\alpha s_0 z} + \sum_{j=1}^{4} ins_j \omega_{1j} \alpha \bar{A}_j e^{\alpha s_j z},$$

$$\tilde{\tilde{D}}_x = ins_0\rho_2\alpha A_0 e^{-\alpha s_0 z} - \sum_{j=1}^{4} ims_j\omega_{2j}\alpha A_j e^{-\alpha s_j z}$$

$$- ins_0\rho_2\alpha\bar{A}_0 e^{\alpha s_0 z} + \sum_{j=1}^{4} ims_j\omega_{2j}\alpha\bar{A}_j e^{\alpha s_j z},$$

$$\tilde{\tilde{D}}_y = -ims_0\rho_2\alpha A_0 e^{-\alpha s_0 z} - \sum_{j=1}^{4} ins_j\omega_{2j}\alpha A_j e^{-\alpha s_j z} \quad \text{(A4)}$$

$$+ ims_0\rho_2\alpha\bar{A}_0 e^{\alpha s_0 z} + \sum_{j=1}^{4} ins_j\omega_{2j}\alpha\bar{A}_j e^{\alpha s_j z},$$

$$\tilde{\tilde{D}}_z = \sum_{j=1}^{4} \omega_{2j}\alpha^2 A_j e^{-\alpha s_j z} + \sum_{j=1}^{4} \omega_{2j}\alpha^2\bar{A}_j e^{\alpha s_j z},$$

$$\tilde{\tilde{B}}_x = ins_0\rho_3\alpha A_0 e^{-\alpha s_0 z} - \sum_{j=1}^{4} ims_j\omega_{3j}\alpha A_j e^{-\alpha s_j z}$$

$$- ins_0\rho_3\alpha\bar{A}_0 e^{\alpha s_0 z} + \sum_{j=1}^{4} ims_j\omega_{3j}\alpha\bar{A}_j e^{\alpha s_j z},$$

$$\tilde{\tilde{B}}_y = -ims_0\rho_3\alpha A_0 e^{-\alpha s_0 z} - \sum_{j=1}^{4} ins_j\omega_{3j}\alpha A_j e^{-\alpha s_j z} \quad \text{(A5)}$$

$$+ ims_0\rho_3\alpha\bar{A}_0 e^{\alpha s_0 z} + \sum_{j=1}^{4} ins_j\omega_{3j}\alpha\bar{A}_j e^{\alpha s_j z},$$

$$\tilde{\tilde{B}}_z = \sum_{j=1}^{4} \omega_{3j}\alpha^2 A_j e^{-\alpha s_j z} + \sum_{j=1}^{4} \omega_{3j}\alpha^2\bar{A}_j e^{\alpha s_j z}.$$

where the coefficients $A_k$ can be obtained are given as follows,

$$A_1 = [q^{(1)}h_2^{(2)} - q^{(2)}h_2^{(1)}] / [h_1^{(1)}h_2^{(2)} - h_2^{(1)}h_1^{(2)}],$$

$$A_2 = [q^{(2)} - h_1^{(2)}A_1] / h_2^{(2)},$$

$$A_3 = [-k_4^{(3)}p^{(2)} / k_4^{(4)} - \sum_{j=1}^{2} l_j^{(3)}A_j] / l_3^{(3)},$$

$$A_4 = [p^{(2)} - \sum_{j=1}^{3} k_j^{(4)}A_j] / k_4^{(4)},$$

$$\bar{A}_1 = -\sum_{j=1}^{4} s_j^{(5)}A_j / \bar{s}_1^{(5)},$$

$$\bar{A}_2 = [-\sum_{j=1}^{4} t_{2j}^{(1)} e^{-\alpha(s_j+s_2)} A_j - \bar{t}_{21}^{(1)} e^{\alpha(s_1-s_2)h_t} \bar{A}_1] / \bar{t}_{22}^{(1)},$$

$$\bar{A}_3 = [-\sum_{j=1}^{4} k_j e^{-\alpha(s_j+s_3)h_t} A_j - \sum_{j=1}^{2} \bar{k}_j e^{\alpha(s_j-s_3)h_t} \bar{A}_j] / \bar{k}_3,$$

$$\bar{A}_4 = -\sum_{j=1}^{4} e^{-\alpha(s_j+s_4)h_t} A_j - \sum_{j=1}^{3} e^{\alpha(s_j-s_4)h_t} \bar{A}_j,$$

$$A_0 = -\sum_{j=1}^{4} n(r_j A_j - \bar{r}_j \bar{A}_j)/(mr_0), \tag{A6}$$

$$\bar{A}_0 = [-\sum_{j=1}^{4} n(e^{-\alpha(s_j+s_4)h_t} A_j - e^{\alpha(s_j-s_4)h_t} \bar{A}_j) - m e^{-\alpha(s_0+s_4)h_t} A_0]/(m e^{\alpha(s_0-s_4)h_t}).$$

The constants in **Eqs. (A1)-(A5)** are given below [10,60]. $s_1, s_2, ..., s_4$ are determined by the four roots of the following equation, all of which have positive real parts.

$$n_0 s^8 - n_1 s^6 + n_2 s^4 - n_3 s^2 + n_4 = 0 \tag{A7}$$

$$n_0 = c_{44}[(c_{33}(\varepsilon_{33}\mu_{33} - g_{33}^2) - 2e_{33}g_{33}d_{33} + \mu_{33}e_{33}^2 + \varepsilon_{33}d_{33}^2], \tag{A8}$$

$$\begin{aligned}
n_1 =\ & c_{11}[c_{33}(\varepsilon_{33}\mu_{33} - g_{33}^2) - 2e_{33}g_{33}d_{33} + \mu_{33}e_{33}^2 + \varepsilon_{33}d_{33}^2] \\
& + c_{44}[c_{44}(\varepsilon_{33}\mu_{33} - g_{33}^2) + c_{33}(\varepsilon_{11}\mu_{33} + \varepsilon_{33}\mu_{11} - 2g_{11}g_{33}) \\
& - 2e_{15}g_{33}d_{33} - 2e_{33}(g_{11}d_{33} + g_{33}d_{15}) + (\mu_{11}e_{33}^2 + 2\mu_{33}e_{15}e_{33}) \\
& + (\varepsilon_{11}d_{33}^2 + 2\varepsilon_{33}d_{15}d_{33})] - (c_{13} + c_{44})[(c_{13} + c_{44})(\varepsilon_{33}\mu_{33} - g_{33}^2) \\
& + (e_{15} + e_{31})(e_{33}\mu_{33} - d_{33}g_{33}) - (d_{15} + d_{31})(e_{33}g_{33} - d_{33}\varepsilon_{33})] \\
& - (e_{15} + e_{31})[(c_{13} + c_{44})(e_{33}\mu_{33} - g_{33}d_{33}) - (e_{15} + e_{31})(c_{33}\mu_{33} + d_{33}^2) \\
& + (d_{15} + d_{31})(c_{33}g_{33} + d_{33}e_{33})] - (d_{15} + d_{31})[(c_{13} + c_{44})(\varepsilon_{33}d_{33} - e_{33}g_{33}) \\
& + (e_{15} + e_{31})(c_{33}g_{33} + e_{33}d_{33}) - (d_{15} + d_{31})(c_{33}\varepsilon_{33} + e_{33}^2)],
\end{aligned} \tag{A9}$$

$$\begin{aligned}
n_2 =\ & c_{11}[c_{44}(\varepsilon_{33}\mu_{33} - g_{33}^2) + c_{33}(\varepsilon_{11}\mu_{33} + \varepsilon_{33}\mu_{11} - 2g_{11}g_{33}) \\
& - 2e_{15}g_{33}d_{33} - 2e_{33}(g_{11}d_{33} + g_{33}d_{15}) + (\mu_{11}e_{33}^2 + 2\mu_{33}e_{15}e_{33}) \\
& + (\varepsilon_{11}d_{33}^2 + 2\varepsilon_{33}d_{15}d_{33})] + c_{44}[c_{44}(\varepsilon_{11}\mu_{33} + \varepsilon_{33}\mu_{11} - 2g_{11}g_{33}) \\
& + c_{33}(\varepsilon_{11}\mu_{11} - g_{11}^2) - 2e_{15}(g_{11}d_{33} + g_{33}d_{15}) \\
& - 2e_{33}g_{11}d_{15} + 2\mu_{11}e_{15}e_{33} + \mu_{33}e_{15}^2 + 2\varepsilon_{11}d_{15}d_{33} + \varepsilon_{33}d_{15}^2] \\
& - (c_{13} + c_{44})[(c_{13} + c_{44})(\varepsilon_{11}\mu_{33} + \varepsilon_{33}\mu_{11} - 2g_{11}g_{33}) \\
& + (e_{15} + e_{31})(e_{15}\mu_{33} + e_{33}\mu_{11}) - (e_{15} + e_{31})(d_{15}g_{33} + d_{33}g_{11}) \\
& - (d_{15} + d_{31})(e_{15}g_{33} + e_{33}g_{11}) + (d_{15} + d_{31})(d_{15}\varepsilon_{33} + d_{33}\varepsilon_{11})] \\
& - (e_{15} + e_{31})[(c_{13} + c_{44})(e_{15}\mu_{33} + e_{33}\mu_{11} - g_{11}d_{33} - g_{33}d_{15}) \\
& - (e_{15} + e_{31})(c_{44}\mu_{33} + c_{33}\mu_{11} + 2d_{15}d_{33}) \\
& + (d_{15} + d_{31})(c_{44}g_{33} + c_{33}g_{11} + d_{15}e_{33} + d_{33}e_{15})] \\
& - (d_{15} + d_{31})[(c_{13} + c_{44})(\varepsilon_{11}d_{33} + \varepsilon_{33}d_{15} - e_{15}g_{33} - e_{33}g_{11}) \\
& + (e_{15} + e_{31})(c_{44}g_{33} + c_{33}g_{11} + e_{15}d_{33} + e_{33}d_{15}) \\
& - (d_{15} + d_{31})(c_{44}\varepsilon_{33} + c_{33}\varepsilon_{11} + 2e_{15}e_{33})],
\end{aligned} \tag{A10}$$

$$n_3 = c_{11}[c_{44}(\varepsilon_{11}\mu_{33} + \varepsilon_{33}\mu_{11} - 2g_{11}g_{33}) + c_{33}(\varepsilon_{11}\mu_{11} - g_{11}^2)$$
$$- 2e_{15}(g_{11}d_{33} + g_{33}d_{15}) - 2e_{33}g_{11}d_{15} + 2\mu_{11}e_{15}e_{33} + \mu_{33}e_{15}^2$$
$$+ 2\varepsilon_{11}d_{15}d_{33} + \varepsilon_{33}d_{15}^2] + c_{44}[c_{44}(\varepsilon_{11}\mu_{11} - g_{11}^2) - 2e_{15}g_{11}d_{15}$$
$$+ \mu_{11}e_{15}^2 + \varepsilon_{11}d_{15}^2] - (c_{13} + c_{44})[(c_{13} + c_{44})(\varepsilon_{11}\mu_{11} - g_{11}^2)$$
$$+ (e_{15} + e_{31})(e_{15}\mu_{11} - d_{15}g_{11}) - (d_{15} + d_{31})(e_{15}g_{11} - d_{15}\varepsilon_{11})]$$
$$- (e_{15} + e_{31})[(c_{13} + c_{44})(e_{15}\mu_{11} - g_{11}d_{15})$$
$$- (e_{15} + e_{31})(c_{44}\mu_{11} + d_{15}^2) + (d_{15} + d_{31})(c_{44}g_{11} + d_{15}e_{15})]$$
$$- (d_{15} + d_{31})[(c_{13} + c_{44})(\varepsilon_{11}d_{15} - e_{15}g_{11})$$
$$+ (e_{15} + e_{31})(c_{44}g_{11} + e_{15}d_{15}) - (d_{15} + d_{31})(c_{44}\varepsilon_{11} + e_{15}^2)],$$
(A11)

$$n_4 = c_{11}[c_{44}(\varepsilon_{11}\mu_{11} - g_{11}^2) - 2e_{15}g_{11}d_{15} + \mu_{11}e_{15}^2 + \varepsilon_{11}d_{15}^2].$$
(A12)

$$k_{ij} = \beta_{ij}/(\alpha_j s_j^2) \quad (i = 1,2,3)$$
$$\omega_{1j} = c_{44}(1 + k_{1j}) + e_{15}k_{2j} + d_{15}k_{3j},$$
$$\omega_{2j} = e_{15}(1 + k_{1j}) - \varepsilon_{11}k_{2j} - g_{11}k_{3j},$$
$$\omega_{3j} = d_{15}(1 + k_{1j}) - g_{11}k_{2j} - \mu_{11}k_{3j},$$
$$\rho_1 = c_{44}, \rho_2 = e_{15}, \rho_3 = d_{15},$$
(A13)

where

$$\alpha_j = -a_1 + a_2 s_j^2 - a_3 s_j^4,$$
$$\beta_{ij} = -a_{4i} + a_{5i}s_j^2 - a_{6i}s_j^4 + a_{7i}s_j^6 \quad (i = 1,2,3),$$
(A14)

with

$$a_1 = (c_{13} + c_{44})(\varepsilon_{11}\mu_{11} - g_{11}^2) + (e_{15} + e_{31})(e_{15}\mu_{11} - g_{11}d_{15})$$
$$- (d_{15} + d_{31})(e_{15}g_{11} - \varepsilon_{11}d_{15}),$$
(A15)

$$a_2 = (c_{13} + c_{44})(\varepsilon_{11}\mu_{33} + \varepsilon_{33}\mu_{11} - 2g_{11}g_{33})$$
$$+ (e_{15} + e_{31})(e_{15}\mu_{33} + e_{33}\mu_{11} - g_{11}d_{33} - g_{33}d_{15})$$
$$- (d_{15} + d_{31})(e_{15}g_{33} + e_{33}g_{11} - \varepsilon_{11}d_{33} - \varepsilon_{33}d_{15}),$$
(A16)

$$a_3 = (c_{13} + c_{44})(\varepsilon_{33}\mu_{33} - g_{33}^2) + (e_{15} + e_{31})(e_{33}\mu_{33} - g_{33}d_{33})$$
$$- (d_{15} + d_{31})(e_{33}g_{33} - \varepsilon_{33}d_{33}),$$
(A17)

$$a_{41} = c_{11}(\varepsilon_{11}\mu_{11} - g_{11}^2),$$
(A18)

$$a_{51} = c_{11}(\varepsilon_{11}\mu_{33} + \varepsilon_{33}\mu_{11} - 2g_{11}g_{33}) + c_{44}(\varepsilon_{11}\mu_{11} - g_{11}^2)$$
$$+ \mu_{11}(e_{15} + e_{31})^2 + \varepsilon_{11}(d_{15} + d_{31})^2$$
$$- 2g_{11}(e_{15} + e_{31})(d_{15} + d_{31}),$$
(A19)

$$a_{61} = c_{11}(\varepsilon_{33}\mu_{33} - g_{33}^2) + c_{44}(\varepsilon_{11}\mu_{33} + \varepsilon_{33}\mu_{11} - 2g_{11}g_{33})$$
$$+ \mu_{33}(e_{15} + e_{31})^2 + \varepsilon_{33}(d_{31} + d_{15})^2$$
$$- 2g_{33}(e_{15} + e_{31})(d_{15} + d_{31}),$$
(A20)

$$a_{71} = c_{44}(\varepsilon_{33}\mu_{33} - g_{33}^2), \tag{A21}$$

$$a_{42} = c_{11}(e_{15}\mu_{11} - g_{11}d_{15}), \tag{A22}$$

$$\begin{aligned}a_{52} = &\, c_{11}(e_{15}\mu_{33} + e_{33}\mu_{11} - g_{11}d_{33} - g_{33}d_{15}) \\ &+ c_{44}(e_{15}\mu_{11} - g_{11}d_{15}) \\ &- (e_{15} + e_{31})[\mu_{11}(c_{13} + c_{44}) + d_{15}(d_{15} + d_{31})] \\ &+ (d_{15} + d_{31})[g_{11}(c_{13} + c_{44}) + e_{15}(d_{15} + d_{31})],\end{aligned} \tag{A23}$$

$$\begin{aligned}a_{62} = &\, c_{11}(e_{33}\mu_{33} - g_{33}d_{33}) \\ &+ c_{44}(e_{15}\mu_{33} + e_{33}\mu_{11} - g_{11}d_{33} - g_{33}d_{15}) \\ &- (e_{15} + e_{31})[\mu_{33}(c_{13} + c_{44}) + d_{33}(d_{15} + d_{31})] \\ &+ (d_{15} + d_{31})[g_{33}(c_{13} + c_{44}) + e_{33}(d_{15} + d_{31})],\end{aligned} \tag{A24}$$

$$a_{72} = c_{44}(e_{33}\mu_{33} - g_{33}d_{33}), \tag{A25}$$

$$a_{43} = c_{11}(\varepsilon_{11}d_{15} - e_{15}g_{11}), \tag{A26}$$

$$\begin{aligned}a_{53} = &\, c_{11}(-e_{15}g_{33} - e_{33}g_{11} + \varepsilon_{11}d_{33} + \varepsilon_{33}d_{15}) \\ &+ c_{44}(-e_{15}g_{11} + \varepsilon_{11}d_{15}) \\ &+ (e_{15} + e_{31})[g_{11}(c_{13} + c_{44}) + d_{15}(e_{15} + e_{31})] \\ &- (d_{15} + d_{31})[\varepsilon_{11}(c_{13} + c_{44}) + e_{15}(e_{15} + e_{31})],\end{aligned} \tag{A27}$$

$$\begin{aligned}a_{63} = &\, c_{11}(-e_{33}g_{33} + \varepsilon_{33}d_{33}) \\ &+ c_{44}(-e_{15}g_{33} - e_{33}g_{11} + \varepsilon_{11}d_{33} + \varepsilon_{33}d_{15}) \\ &+ (e_{15} + e_{31})[g_{33}(c_{13} + c_{44}) + d_{33}(e_{15} + e_{31})] \\ &- (d_{15} + d_{31})[\varepsilon_{33}(c_{13} + c_{44}) + e_{33}(e_{15} + e_{31})],\end{aligned} \tag{A28}$$

$$a_{73} = c_{44}(-e_{33}g_{33} + \varepsilon_{33}d_{33}). \tag{A29}$$

The intermediate variables in **Eq. (A5)** are:

$$\begin{aligned}&t_{1j} = \omega_{1j}, \quad \overline{t}_{1j} = \omega_{1j}, \\ &t_{2j} = -s_j\omega_{1j}, \\ &\overline{t}_{2j} = s_j\omega_{1j}, \\ &w_{2j} = \omega_{2j}, \\ &\overline{w}_{2j} = \omega_{2j}, \\ &w_{3j} = \omega_{3j}, \\ &\overline{w}_{3j} = \omega_{3j}.\end{aligned} \tag{A30}$$

$$\begin{aligned}&r_j = -t_{2j} + \rho_1 s_0 e^{-\alpha(s_j + s_0)h_t}, \\ &\overline{r}_j = -\overline{t}_{2j} + \rho_1 s_0 e^{\alpha(s_j - s_0)h_t}, \\ &r_0 = \rho_1 s_0 + \rho_1 s_0 e^{-2\alpha s_0 h_t}, \\ &r_{ij} = s_j(k_{ij} - k_{i4}k_{3j}/k_{34}).\end{aligned} \tag{A31}$$

$$k_j = s_4 + s_j k_{3j} / k_{34}, \tag{A32}$$
$$\bar{k}_j = s_4 - s_j k_{3j} / k_{34}.$$

$$\begin{aligned} t_j^{(1)} &= t_{1j} - \bar{t}_{14} t_{2j} / \bar{t}_{24}, & t_j^{(2)} &= w_{2j} - \bar{w}_{24} w_{3j} / \bar{w}_{34}, \\ \bar{t}_j^{(1)} &= \bar{t}_{1j} - \bar{t}_{14} \bar{t}_{2j} / \bar{t}_{24}, & \bar{t}_j^{(2)} &= \bar{w}_{2j} - \bar{w}_{24} \bar{w}_{3j} / \bar{w}_{34}, \\ t_j^{(3)} &= w_{3j} - \bar{w}_{34} e^{-\alpha(s_j + s_4) h_t}, & t_j^{(4)} &= t_{2j} - \bar{t}_{24} e^{-\alpha(s_j + s_4) h_t}, \\ \bar{t}_j^{(3)} &= \bar{w}_{3j} - \bar{w}_{34} e^{\alpha(s_j - s_4) h_t}, & \bar{t}_j^{(4)} &= \bar{t}_{2j} - \bar{t}_{24} e^{\alpha(s_j - s_4) h_t}. \end{aligned} \tag{A33}$$

$$\begin{aligned} t_{ij}^{(1)} &= r_{ij} \bar{k}_3 / r_{i3} + k_j, \\ \bar{t}_{ij}^{(1)} &= -r_{ij} \bar{k}_3 / r_{i3} + \bar{k}_j, \\ r_j^{(i)} &= t_j^{(i)} \bar{k}_3 / \bar{t}_3^{(i)} - k_j e^{-\alpha(s_j + s_3) h_t}, \\ \bar{r}_j^{(i)} &= \bar{t}_j^{(i)} \bar{k}_3 / \bar{t}_3^{(i)} - \bar{k}_j e^{\alpha(s_j - s_3) h_t}, \\ s_j^{(i)} &= r_j^{(i)} - \bar{r}_2^{(i)} t_{2j}^{(1)} e^{-\alpha(s_j + s_2) h_t} / \bar{t}_{22}^{(1)}, \\ \bar{s}_1^{(i)} &= \bar{r}_1^{(i)} - \bar{r}_2^{(i)} \bar{t}_{21}^{(1)} e^{\alpha(s_1 - s_2) h_t} / \bar{t}_{22}^{(1)}, \\ s_j^{(5)} &= (t_{1j}^{(1)} - \bar{t}_{12}^{(1)} t_{2j}^{(1)} / \bar{t}_{22}^{(1)}) e^{-\alpha(s_j + s_1) h_t}, \\ \bar{s}_1^{(5)} &= (\bar{t}_{11}^{(1)} - \bar{t}_{12}^{(1)} \bar{t}_{21}^{(1)} / \bar{t}_{22}^{(1)}), \\ k_j^{(i)} &= s_j^{(i)} - \bar{s}_1^{(i)} s_j^{(5)} / \bar{s}_1^{(5)}. \end{aligned} \tag{A34}$$

$$\begin{aligned} p^{(1)} &= (-\tilde{\tilde{p}} / \alpha^2 - im\mu_f \bar{t}_{14} \tilde{\tilde{p}} / \bar{t}_{24} / \alpha^3) \bar{k}_3 / \bar{t}_3^{(1)}, \\ p^{(2)} &= [\bar{w}_{24} * \tilde{\tilde{g}}_b / (\bar{w}_{34} \alpha^2) - \tilde{\tilde{q}}_b / \alpha^2] \bar{k}_3 / \bar{t}_3^{(2)}, \\ p^{(3)} &= (-\tilde{\tilde{g}}_b / \alpha^2) \bar{k}_3 * \bar{t}_{22}^{(1)} / \bar{t}_3^{(3)}, \\ p^{(4)} &= (im\mu_f \tilde{\tilde{p}} / \alpha^3) \bar{k}_3 / \bar{t}_3^{(4)}. \end{aligned} \tag{A35}$$

$$\begin{aligned} l_j^{(i)} &= k_j^{(i)} - k_4^{(i)} k_j^{(4)} / k_4^{(4)}, \\ h_j^{(i)} &= l_j^{(i)} - l_3^{(i)} l_j^{(3)} / l_3^{(3)}. \end{aligned} \tag{A36}$$

$$\begin{aligned} q^{(1)} &= (p^{(1)} - k_4^{(1)} p^{(4)} / k_4^{(4)}) - (p^{(3)} - k_4^{(3)} p^{(4)} / k_4^{(4)}) * l_3^{(1)} / l_3^{(3)}, \\ q^{(2)} &= (p^{(2)} - k_4^{(2)} p^{(4)} / k_4^{(4)}) - (p^{(3)} - k_4^{(3)} p^{(4)} / k_4^{(4)}) * l_3^{(2)} / l_3^{(3)}. \end{aligned} \tag{A37}$$